\documentclass[12pt,letter]{article}
\usepackage{graphicx,subfigure}
\usepackage{epsfig,amssymb}
\newcommand{\rom}[1]{\mathrm{#1}}

\newcommand{\beq}{\begin{equation}}
\newcommand{\eeq}{\end{equation}}
\newcommand{\be}{\begin{equation}}
\newcommand{\ee}{\end{equation}}
\newcommand{\beqa}{\begin{eqnarray}}
\newcommand{\eeqa}{\end{eqnarray}}
\newcommand{\beqar}{\begin{eqnarray*}}
\newcommand{\eeqar}{\end{eqnarray*}}
\newcommand{\bea}{\begin{eqnarray*}}
\newcommand{\eea}{\end{eqnarray*}}

\newcommand{\p}{\partial}

\newcommand{\reef}[1]{(\ref{#1})}

\newcommand{\eg}{{\it e.g.,}\ }
\newcommand{\ie}{{\it i.e.,}\ }

\newcommand{\gsim}{\mathrel{\raisebox{-.6ex}{$\stackrel{\textstyle>}{\sim}$}}}

\begin{document}

\setlength{\unitlength}{1mm}

\begin{titlepage}

\begin{flushright}
hep-th/0608076\\
MIT-CTP-3760\\
NSF-KITP-06-50
\end{flushright}
\vspace{1cm}

\begin{center}
{\bf \Large Dynamics and Stability of Black Rings}
\end{center}

\vspace{1cm}

\begin{center}
Henriette Elvang$^{a}$,  Roberto Emparan$^{b}$, and Amitabh
Virmani$^{c}$

\vspace{.5cm}
{\small {\textit{$^{a}$Center for Theoretical Physics, }}\\
{\small\textit{Massachusetts
Institute of Technology, Cambridge, MA 02139, USA}}}\\
\vspace{2mm}
{\small \textit{$^{b}$Instituci\'o Catalana de Recerca i Estudis
Avan\c cats (ICREA)}}\\
{\small\textit{and}}\\
{\small\textit{Departament de F{\'\i}sica Fonamental}}\\
{\small\textit{Universitat de
Barcelona, Diagonal 647, E-08028, Barcelona, Spain}}\\
\vspace{2mm}
{\small {\textit{$^{c}$Department of Physics, University of
California, Santa Barbara, CA 93106-9530, USA}}}

\vspace*{0.5cm}
{\tt elvang@lns.mit.edu, emparan@ub.edu, virmani@physics.ucsb.edu}
\end{center}

\vspace{1cm}

\begin{abstract}

We examine the dynamics of neutral black rings, and identify and analyze
a selection of possible instabilities. We find the dominating forces of
very thin black rings to be a Newtonian competition between a
string-like tension and a centrifugal force. We study in detail the
radial balance of forces in black rings, and find evidence that all fat
black rings are unstable to radial perturbations, while thin black rings
are radially stable. Most thin black rings, if not all of them, also
likely suffer from Gregory-Laflamme instabilities. We also study simple
models for stability against emission/absorption of massless particles.
Our results point to the conclusion that most neutral black rings suffer
from classical dynamical instabilities, but there may still exist a
small range of parameters where thin black rings are stable. We also
discuss the absence of regular real Euclidean sections of black rings,
and thermodynamics in the grand-canonical ensemble.

\end{abstract}

\end{titlepage}

\vspace{0.5cm}
\tableofcontents

\newpage

\setcounter{equation}{0}
\section{Introduction and Summary of Results}
\label{sec:intro}

Black holes are much more interesting than what their image as spacetime
sinks suggests. They are highly dynamical objects that can vibrate and
pulsate, react to external action and interact with their environment. Thus
black holes become rather similar to objects in other realms of physics
--- but still they are made only of curved spacetime.

A major tool in
understanding the dynamics of four-dimensional black holes is the
ordinary differential equation, derived by Teukolsky, that encodes the
dynamics of linearized perturbations, in particular tensorial ones,
around the Kerr solution \cite{chandrabook}. The linear stability of the Kerr
black hole is a main spinoff of this equation.

In higher dimensions, the larger number of degrees of freedom naturally
gives gravity richer dynamics. A dramatic manifestation of the new
possibilities is the existence of five-dimensional black rings, which
bring in non-spherical horizon topologies and the absence of black hole
uniqueness \cite{ER} (see \cite{review} for a recent review). However,
this richness also implies a much greater complexity in the equations
governing their perturbations, and so, the five-dimensional analogue of
Teukolsky's ordinary differential equation, if it exists at all, has
remained elusive so far for both the black rings and the topologically
spherical Myers-Perry (MP) black holes \cite{MP} (see \cite{KLR} for
recent results). Since the problems appear to be of a technical rather
than conceptual nature, numerical methods are presumably useful here.

For the time being, some qualitative and semi-quantitative
analytical approaches have been advanced (\eg \cite{ER, grf, EM,
loar,Maeda, Oscar,HovMyers}), and in this paper we provide further
progress in this direction. As was the case with the analysis of the
dynamics of rotating MP black holes in $D>5$ in \cite{EM}, we hope
that our results help to guide future studies of the instabilities
of black rings.

Our purpose is to examine the dynamics of neutral black rings, and to
identify and analyze a selection of possible instabilities. The methods
involve approximations that are often crude, but we devote quite some
effort to identify how far they can be reliably pushed. Putting things
together, our results point to the conclusion that, over a very wide
parameter range, neutral black rings suffer from classical
instabilities, although for a small range of parameters close to the
regime of non-uniqueness, thin black rings may be stable. Besides this,
we also perform fairly detailed studies that have been lacking in the
literature so far. We discuss the geometry of the neutral black ring ---
shapes and distortions --- and the mechanical balance of forces that the
black ring embodies. We consider possible real Euclidean sections of the
black ring and show that they fail to be non-singular, and we discuss 5D
black hole thermodynamics in the grand-canonical ensemble.

\begin{figure}[t]
\begin{picture}(0,0)(0,0)
\put(102.5, -3){$\scriptscriptstyle{1}$}
\put(91,-3){$\scriptscriptstyle{\frac{27}{32}}$}
\put(34,21){$\scriptscriptstyle{1}$}
\put(31,60){$\scriptscriptstyle{2\sqrt{2}}$}
\put(139,-1){$\scriptstyle{j^2}$}
\put(35,65){$\scriptstyle{a_\rom{H}}$}
\put(114,19){{\footnotesize thin black ring}}
\put(76,9){{\footnotesize fat black ring}}
\put(73,47){{\footnotesize MP black hole}}
\end{picture}
\vspace{2mm}
\centerline{
\includegraphics[width=10cm]{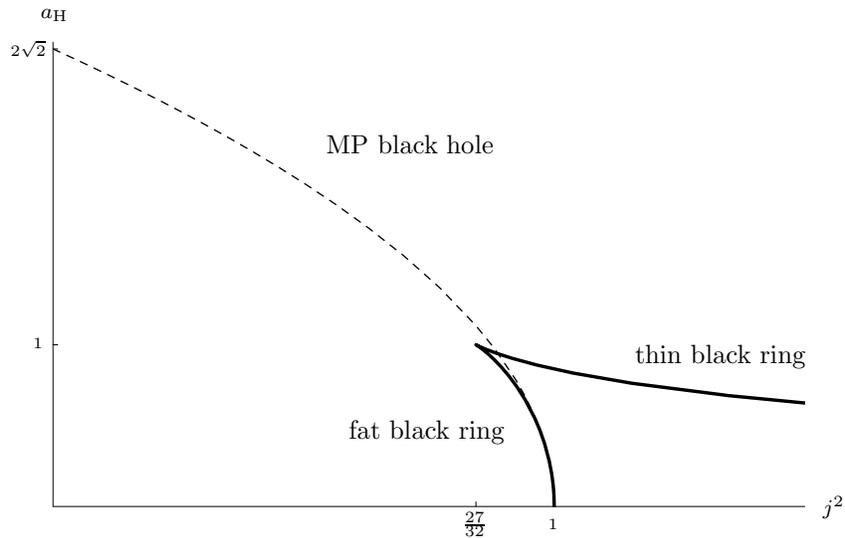}}
\caption{\small Phase diagram of the five-dimensional black rings (solid)
and MP black holes (dotted). We plot area (\ie entropy) vs.\
angular momentum $j^2$, both in reduced units for fixed mass.}
\label{fig:phase}
\end{figure}

Up to an overall scale, the black rings of \cite{ER} and the MP black holes
are conveniently parametrized by a single dimensionless
parameter $j$, which represents the angular momentum for fixed mass.
The phase diagram showing entropy versus $j$ for fixed mass
has been discussed at length earlier \cite{ER,review,EE,RE},
and we reproduce it in fig.~\ref{fig:phase} for reference.
When discussing black rings, we conveniently distinguish between the
``thin black ring branch'' (dominating entropically) and the ``fat black
ring branch'', see fig.~\ref{fig:phase}.

\vspace{4mm}
\noindent Our results can be summarized in simple physical terms:

\begin{itemize}

\item {\bf The shape of black rings (section \ref{sec:neutral})}:
\begin{itemize}
\item A solution on
the thin black ring branch has a horizon with an $S^2$ that is nearly
round. On the fat black ring branch, horizons are highly distorted and,
as they approach the singular limit $j \to 1$, they flatten out along the rotation
plane, with the inner ring circle shrinking to zero and the outer circle
becoming infinitely long.

\end{itemize}

\item {\bf Black Ring Mechanics (section \ref{sec:balance})}:
\begin{itemize}
\item
We study the physics of the equilibrium condition for black rings in the
large-$j$ limit, \ie very thin black rings. To leading order, the rings
are balanced by a competition between a string-like tension and a
centrifugal force. In fact, the balance of forces can be reproduced
exactly by a Newtonian calculation.
The gravitational attractive force appears only at
subleading order. Hence the equilibrium of thin black rings is
independent of the number of dimensions and can be expected also in
$D>5$.
\end{itemize}

\item {\bf Radial stability and the radial potential (sections
\ref{sec:radial-general} and \ref{sec:turning})}:
\begin{itemize}
\item
We consider a family of off-shell black rings radially deformed away
from equilibrium. This allows us to decide whether the black ring sits
at a maximum or a minimum of a radial potential $V(R)$. We find that fat
black rings are always at a maximum (hence radially unstable) while thin
black rings sit at minima (hence radially stable). The black ring with
minimal spin sits at an inflection point and so is also unstable to
radial perturbations. 
We argue that the radial instability is the
classical instability that ref.~\cite{loar} argued to exist for fat black
rings but not for thin black rings. 

\end{itemize}

\item {\bf Gregory-Laflamme instability and fragmentation (section
\ref{sec:gl}):}
The Gregory-La\-flamme (GL) instability \cite{GL1,GL2} of thin black
rings to development of inhomogeneities along the circle of the ring was suggested
early on \cite{ER}, and was then subjected to closer study in \cite{HovMyers}.
We try to push the study to thicker black rings by using our detailed
analysis of the properties of the ring.

\begin{itemize}

\item It is possible to fit several GL instability modes on all thin black rings.
Thus we recover the expectation \cite{ER,HovMyers} that thin black
rings with large $j$ are unstable to Gregory-Laflamme type
instabilities, and suggest that this may extend down to moderate values
$j\gsim O(1)$.

\item
It is always entropically favorable for a black ring to fragment into two
or more spherical black holes. This motivates a possible endstate for the GL instability. 

\end{itemize}

\item {\bf Instabilities via emission/capture of null particles (section
\ref{sec:geodesics}):}
\begin{itemize}
\item Absorption of a counter-rotating null-particle by the black ring
is always possible, and $j$ necessarily decreases in the process. This
makes it possible to underspin the minimal-$j$ black ring. Since there
are no black ring solutions with smaller $j$, this indicates that the
slowest-spinning black
ring {\it is unstable} and will most likely collapse to a MP black hole.
\item One cannot overspin the MP black hole nor fat black rings (and
thus violate cosmic censorship), by throwing in a co-rotating null
particle.
\item It is not possible for the MP black hole nor for the black ring to spontaneously
decay by emitting null particles carrying away energy and angular
momentum without violating the area law.
\end{itemize}

\item {\bf Thermodynamics (section \ref{sec:thermo})}:
\begin{itemize}
\item
  Real Euclidean metrics obtained by analytic
  continuation of the black ring suffer from either conical or naked
  singularities. Thus there is no candidate metric for a real
  Euclidean section of the neutral black ring.
\item We define a grand-canonical potential
  $W=W(T,\Omega)$. For given $\Omega$, there exist for any $T>0$
  one Myers-Perry black hole and one black ring. Since $W_\rom{MP} <
  W_\rom{BR}$, in the grand canonical ensemble the MP black hole is
  the thermodynamically favored stable solution.
\end{itemize}

\end{itemize}

In the final section \ref{sec:discussion}, we argue that these results
point to the conclusion that neutral black rings can only be dynamically
stable, if at all, over a narrow range of parameters, and we discuss the
consequences.


\setcounter{equation}{0}
\section{The Geometry of the Neutral Black Ring}
\label{sec:neutral}

\subsection{Metric and properties}
The neutral black ring has been analyzed in detail elsewhere
\cite{ER,EE,RE,review}. For the purpose of reference, we collect
here the results we need. The metric is \cite{RE}\footnote{Following \cite{review},
the sense of rotation of the metric (\ref{neutral}) is positive, hence reversed
relative to \cite{RE}.}

\beqa
  \label{neutral}
  ds^2 &=& -\frac{F(y)}{F(x)} \left( dt-C  \,  R\, \frac{1+y}{F(y)} d\psi\right)^2 \nonumber \\[1mm]
       && + \frac{R^2}{(x-y)^2} F(x)
       \left[
         -\frac{G(y)}{F(y)} d\psi^2
         -\frac{dy^2}{G(y)}
         +\frac{dx^2}{G(x)}
         +\frac{G(x)}{F(x)} d\phi^2
       \right] \, ,
\eeqa
where
\be
  F(\xi) = 1+\lambda \,\xi \, ,
  \qquad
  G(\xi) = (1-\xi^2)(1+\nu \xi) \, ,
  \qquad
  C  = \sqrt{\lambda(\lambda-\nu)\frac{1+\lambda}{1-\lambda}}\,.
\ee
The parameters $\lambda$ and $\nu$ take values
$0< \nu \le \lambda <1$.
The coordinate ranges are $-1 \le x \le 1$, $-\infty < y  \le -1$,
with asymptotic infinity at $x,y=-1$.
Regularity at infinity requires the angular coordinates to have periodicities
\be\label{afperiod}
  \Delta\psi = \Delta\phi = 2\pi \frac{\sqrt{1-\lambda}}{1-\nu}\,.
\ee
Balancing the ring fixes the parameter $\lambda$ in terms of $\nu$ as
\be
  \label{balance}
  \lambda=\lambda_c \equiv \frac{2\nu}{1+\nu^2} \, ,
\ee
which follows from requiring the absence of a conical singularity at $x=1$.

The ergosurface is located at $y=-1/\lambda$ and the event horizon
is at $y=-1/\nu$; both have topology $S^1 \times S^2$. The metric on
a spatial cross-section of the horizon can be written \be
  \label{horizon}
  ds_\rom{H}^2 = R^2\left(\frac{\nu^2 (1+\lambda\, x)}{(1+\nu\, x)^3}
\frac{dx^2}{(1-x^2)}
  + \frac{\nu^2 (1-x^2)}{1+\nu\,x}\, d\phi^2
+\frac{\lambda (1+\lambda)(1-\nu)^2}{\nu (1-\lambda) (1+\lambda \,
x)}\, d\psi^2\right) \, .
\ee
The physical parameters for the black ring, mass, angular
momentum, temperature, angular velocity and horizon area are \beqa
    \label{M}
    M &=& \frac{3\pi R^2}{4 G} \frac{\lambda}{1-\nu} \, , \\[1mm]
    \label{J}
J &=& \frac{\pi R^3}{2 G}
\frac{\sqrt{\lambda(\lambda-\nu)(1+\lambda)}}{(1-\nu)^2} \, , \\[1mm]
T &=& \frac{1}{4\pi R}
\frac{1+\nu}{\sqrt{\nu}}\sqrt{\frac{1-\lambda}{\lambda(1+\lambda)}} \, ,
\\[1mm]
    \label{Omega}
\Omega &=& \frac{1}{R} \sqrt{\frac{\lambda-\nu}{\lambda (1+\lambda)}} \,
, \\[1mm]
   {\cal A_\rom{H}} &=& 8 \pi^2 R^3 \frac{\nu^{3/2}\sqrt{\lambda(1-\lambda^2)}}{(1-\nu)^2(1+\nu)}\,.
    \label{AH}
\eeqa
We fix a scale by fixing the mass and introduce the dimensionless `reduced'
angular momentum and horizon area as
\beqa
   \label{jandaH}
   j = \sqrt{\frac{27 \pi}{32G}} \frac{J}{M^{3/2}} \, ,
   ~~~~~~~
   a_\rom{H} = \frac{3}{16} \sqrt{\frac{3}{\pi}} \frac{{\cal A_\rom{H}}}{(GM)^{3/2}} \, .
\eeqa
For balanced black rings (with $\lambda$ fixed according to
\reef{balance}) one finds a parametric relation between them in the form
\beqa
  \label{jandaH2}
  j^2 = \frac{(1+\nu)^3}{8 \nu} \, ,
  ~~~~~~~
  a_\rom{H} = 2\sqrt{\nu(1-\nu)}\, .
\eeqa
Figure \ref{fig:phase} shows $a_\rom{H}$ versus $j$. The plot
   illustrates the non-uniqueness of black rings and the Myers-Perry
   black hole. The cusp is located at $\nu=1/2$, where $j=\sqrt{27/32}$.
Based on the plot it is natural to distinguish between the two branches
of black rings:
\begin{itemize}
\item {\sl Thin black rings} with $0<\nu<1/2$. This branch extends to
arbitrarily large $j$ as $\nu\to 0$ (`very thin' rings).
\item {\sl Fat black rings} with $1/2<\nu<1$. This branch terminates at
the singular solution with $\nu=1$.
\end{itemize}
Note that $\nu$ can be thought of as a `shape parameter', giving a measure of the
thickness of the ring.

\begin{figure}
\begin{picture}(0,0)(0,0)
\put(44.7, -3){$\scriptscriptstyle{1}$}
\put(36.5,-4){$\scriptscriptstyle{\sqrt{\frac{27}{32}}}$}
\put(127.5,-3){$\scriptscriptstyle{1}$}
\put(119,-4){$\scriptscriptstyle{\sqrt{\frac{27}{32}}}$}
\put(0,12){$\scriptscriptstyle{\frac{1}{4}\sqrt{\frac{3}{2\pi}}}$}
\put(82.5,42){$\scriptscriptstyle{\frac{1}{2}\sqrt{\frac{3\pi}{2}}}$}
\put(83,-1){$\scriptstyle{j}$}
\put(79.5,-3){$\scriptscriptstyle{2}$}
\put(166,-1){$\scriptstyle{j}$}
\put(162.3,-3){$\scriptscriptstyle{2}$}
\put(5,48){$\scriptstyle{T\sqrt{G\, M}}$}
\put(87,48){$\scriptstyle{\Omega \sqrt{G\, M}}$}
\end{picture}
\vspace{2mm}
\centerline{
\includegraphics[width=7cm]{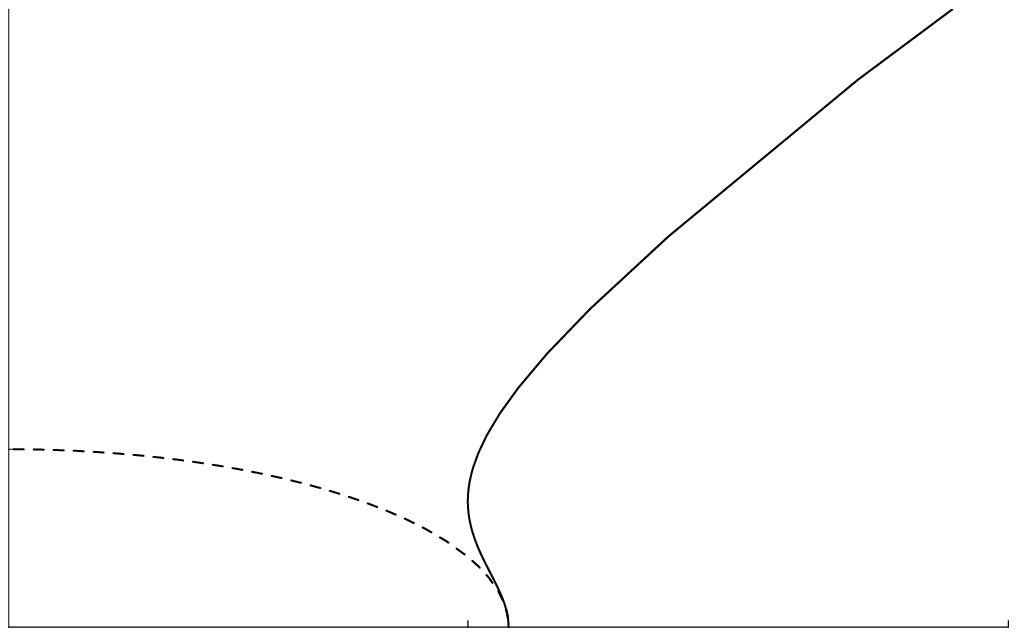}
\hspace{1cm}
\includegraphics[width=7cm]{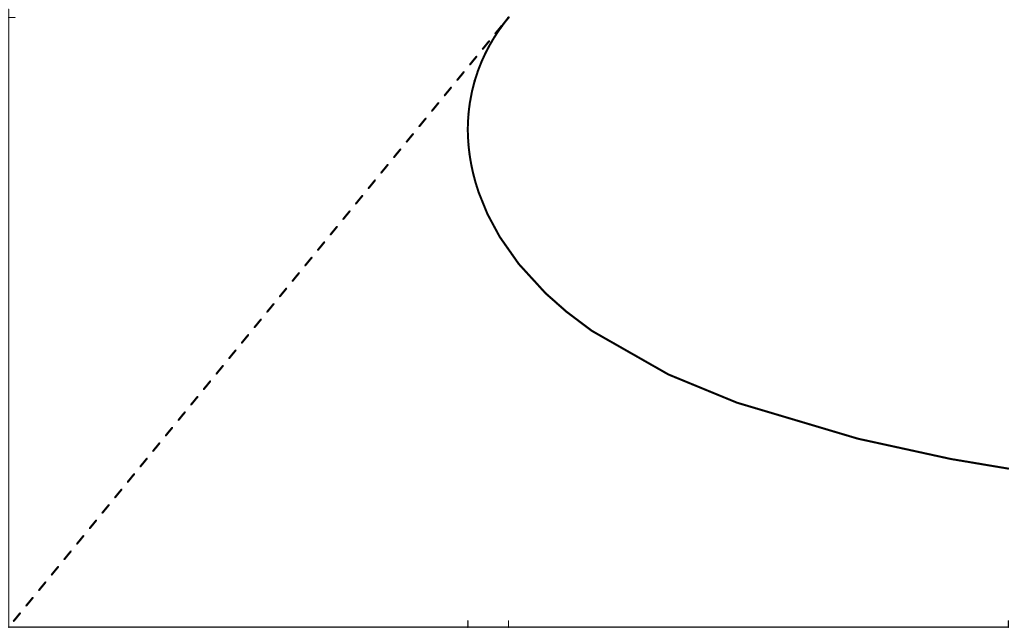}}
\caption{\small Temperature (left) and angular velocity (right) vs.\
angular momentum  $j$ for black rings (solid) and MP black holes (dotted) of
fixed mass. Both $T$ and $\Omega$ have been rendered dimensionless by
multiplying by $\sqrt{GM}$.}
\label{fig:double}
\end{figure}

Figure \ref{fig:double} shows how the temperature and angular velocity
change with $j$. In the regime of non-uniqueness, thin black rings are
the hottest and slowest rotating objects, and MP black holes the coldest
and most rapidly rotating ones. It is curious that, as the angular
momentum of thin black rings grows, their angular velocity decreases and
their temperature increases, in contrast to MP black holes. Both effects
are actually easily understood. The angular velocity decreases because
as we increase the ring radius, the rotation velocity required to maintain $J$
fixed (at constant $M$) becomes smaller. The increasing temperature is a
consequence of the fact that a thin ring is roughly like a circular
black string, and the temperature of black strings is inversely proportional to the
radius of the $S^2$. This radius shrinks as the ring gets thinner, so
its temperature grows.


\subsection{Shape}
\label{sec:shape}
\begin{figure}[t!]
\begin{picture}(0,0)(0,0)
\put(12,13){$\scriptscriptstyle{\nu = 0.05}$}
\put(85,22){$\scriptscriptstyle{\nu = 0.5}$}
\put(125,23){$\scriptscriptstyle{\nu = 0.95}$}
\end{picture}
\centerline{
\includegraphics[width=10cm]{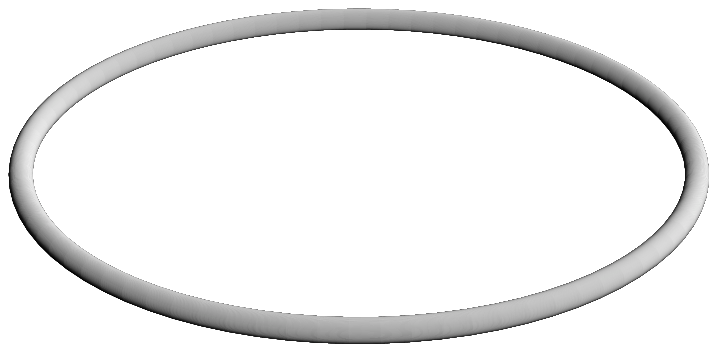}
\hspace{-4.4cm}
\includegraphics[width=10cm]{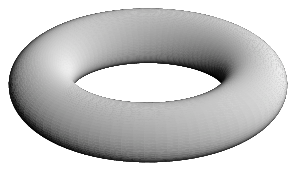}
\hspace{-6.7cm}
\includegraphics[width=10cm]{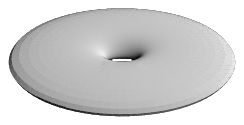}}
\vspace{-1.2cm}
\caption{{\small  Black ring visuals. The azimuthal angle of the $S^2$
    is suppressed. The plot shows the isometric embedding of the $S^2$
    cross section (see fig.~\ref{fig:S2dist}) with the size of the
$S^1$ estimated as the inner radius of the horizon. All three rings
shown have the same mass, and the
    black rings with $\nu=0.05$ and $\nu=0.95$ also have the same
    horizon area.}
\vspace{0.5cm}}
\label{fig:ringsillustrated}
\end{figure}

The topology of the horizon is $S^1 \times S^2$, but metrically
the geometry is not a simple product of an $S^1$ and a round $S^2$.
Fig.~\ref{fig:ringsillustrated} gives an idea of how much rings with
the same mass can vary in shape. In this subsection we study in detail
the shape of black rings, and we introduce measures for
the radii that characterize the $S^1$ and $S^2$ of the
horizon.

\begin{figure}[t!]
\begin{picture}(0,0)(0,0)
\put(8,29){{\small outside}}
\put(152,29){{\small inside}}
\put(22,33){$\scriptstyle{\nu=0.995}$}
\put(38.5,39){$\scriptstyle{\nu=0.95}$}
\put(68,50){$\scriptstyle{\nu=0.8}$}
\put(101,59){$\scriptstyle{\nu=0.5}$}
\put(93.5,27){$\scriptstyle{\nu=0.2}$}
\put(106,29){$\scriptstyle{\nu=0.05}$}
\put(118,30){$\scriptscriptstyle{0.005}$}
\end{picture}
\centerline{
\includegraphics[width=13cm]{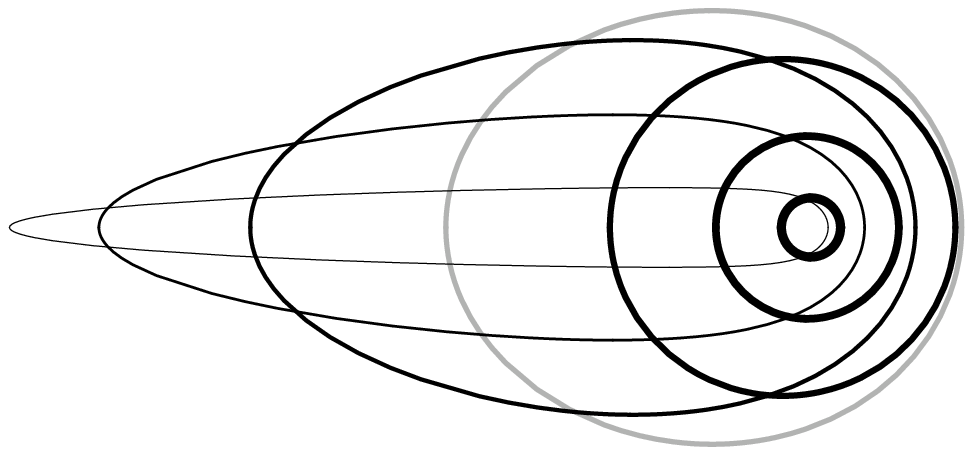}}
\begin{eqnarray*}
{\small
~~~~~
\begin{array}{r|r|r|r|r|r|r|r}

  \nu & 0.005 & 0.05 & 0.2 & 0.5 & 0.8 & 0.95 & 0.995  \\
  \hline
  j   & 5.04  & 1.70 & 1.04& 0.92 & 0.95 & 0.988 & 0.999 \\
\end{array}
}
\end{eqnarray*}
\caption{\small Isometric embedding of the cross-section of the black ring
two-sphere for fixed mass $G M =1$. The azimuthal angle is suppressed. The table shows
values of the reduced angular momentum $j$ for each $\nu$.
The black ring with minimal spin $j^2 = 27/32$ ($\nu=1/2$)
is shown in gray. Solutions on the thin black ring branch ($\nu<1/2$)
have nearly round two-spheres.
The spheres flatten out on the fat black ring branch ($1/2 < \nu < 1$)
as $j \to 1$.
A black ring with $\nu=\nu_0$ has the same horizon area as one with
$\nu=1-\nu_0$.\vspace{0.5cm}}
\label{fig:S2dist}
\end{figure}

The $S^2$ can be highly distorted away from a round sphere, so that
measures of the two-sphere radius convey little information
about the actual geometry.
A useful way of visualizing the shape of the ring, in particular the
distortion of the $S^2$, is via an isometric embedding of the
cross-section of the $S^2$ of the black ring horizon (see
app.~\ref{app:embed}). This is shown in fig.~\ref{fig:S2dist}. We show 
black rings in pairs of equal horizon area: note that $a_\rom{H}$ in
\reef{jandaH2} is invariant under $\nu \to 1- \nu$.

The distortion is particularly important for black rings on the fat
ring branch.
When $j$ approaches $1$ on the fat ring branch ($\nu \to 1$), the
$S^2$ flattens out (for fixed mass) and at $j=1$ the horizon
disappears and the solution is a singular ring. As $j \to 1$ for the
MP black hole, its $S^3$ horizon also flattens out. At $j=1$ the (fat)
black ring and the MP black hole become the same singular solution.

The radii of the ring circle, $R_1$, and two-sphere, $R_2$, on the
horizon are unambiguously defined only for very thin rings
($\lambda,\nu\to 0$), for which $R_1\to R\sqrt{\lambda/\nu}$ and $R_2\to
R\nu$ \cite{RE,review}. In general the $S^2$ is distorted. Furthermore
the size of the $S^1$ depends non-trivially on the polar angle $x \sim
\cos\theta$ of the two-sphere. Several reasonable ways of characterizing
$R_1$ and $R_2$ are possible, see fig.~\ref{fig:sketch} for a sketch, and
appendix~\ref{app:radii} for precise definitions. Salient features of their
behavior
(for black rings of fixed mass) are:

\begin{figure}[t!]
\begin{picture}(0,0)(0,0)
\put(107,22){$\scriptstyle{R_2^\circ}$}
\put(90,10.3){$\scriptstyle{R_1^\circ}$}
\put(84,16){$\scriptstyle{R_1^\rom{inner}}$}
\put(112,1.2){$\scriptstyle{R_1^\rom{outer}}$}
\end{picture}
\centerline{
 \includegraphics[width=8cm]{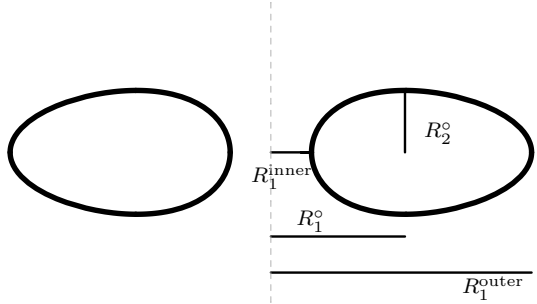}}
 \caption{{\small  Sketch of a diametrical cross section of the black
 ring horizon showing some definitions of radii
 characterizing the black ring horizon: the `equator' radius
 $R_2^\circ$ of the $S^2$, the inner and outer $S^1$ radii,
 $R_1^\rom{inner}$ and $R_1^\rom{outer}$, and finally $R_1^\circ$
 being the radius of the $S^1$ evaluated at the `equator' of the $S^2$.}}\label
{fig:sketch}
\end{figure}

\begin{itemize}

\item
The inner circle radius of the ring, $R_1^\mathrm{inner}$,
decreases monotonically from the thin branch to the fat branch, and
shrinks to zero at the limiting singular solution. The same behavior is
shown by $R_1^\circ$.

\item
The outer circle radius of the ring,  $R_1^\mathrm{outer}$,
decreases along the thin branch until, close to the cusp between
branches, it begins to grow. It then increases monotonically to diverge
at the limiting singular solution.

\end{itemize}

Finally, we also note that the rotational velocity of the black ring results in
relativistic Lorentz effects. For instance, in the limit of a black
string there is a boost factor $\sqrt{\lambda/\nu}$ between the lengths
measured at the horizon and the lengths measured at a large transverse
distance \cite{EE,RE,review}. We discuss in appendix \ref{app:radii} how
to account for this effect for generic rings. Defining a boost velocity
$v$ is mostly sensible for black rings in the thin branch, and the most
important result is that in this branch $v$ always remains moderate and
not too far from its value at infinite radius, $v\to 1/\sqrt{2}$.


\setcounter{equation}{0}
\section{Off-shell Probes of Radial Equilibrium}
\label{sec:radial}

In order to better understand the conditions for equilibrium in a system
it is often useful to study the effect of introducing external
forces, \ie probing the system off-shell. By taking it away from
equilibrium one can obtain information about the potential that the
equilibrium configuration extremizes.

In the case of black rings it is very easy to study the effect of
an external radial force that preserves all the Killing
symmetries of the system but which deforms the radius of the black ring
away from its value at equilibrium. Simply drop the equilibrium
condition \reef{balance}, while still
preserving \reef{afperiod} for asymptotic flatness. Then $\lambda$ is
regarded as a parameter independent of $\nu$.
In general, this creates a
conical defect $\delta$ in the disk inside the ring. This `conical
membrane' creates a tension $\tau$ acting per
unit length of the black ring circle,
\beq\label{taudelta}
  \tau=\frac{3}{16\pi G}\;\delta=
  \frac{3}{8 G}\left(1-\frac{1+\nu}{1-\nu}\sqrt{\frac{1-\lambda}{1+\lambda}}
\right)\, ,
\eeq
which we can view as the force that keeps the
off-balance system in a stationary configuration. The normalization
factor in the definition of $\tau$
will be justified below. Absence of the conical membrane, $\tau=0$, is
equivalent to the equilibrium condition \reef{balance}.

\begin{figure}
\begin{picture}(0,0)(0,0)
\put(50,35){$\scriptstyle{\Delta R_1 > 0}$}
\put(116,37){$\textstyle{\tau < 0}$} \put(116,7){$\textstyle{\tau >
0}$}
\end{picture}
\centerline{
\includegraphics[width=14cm]{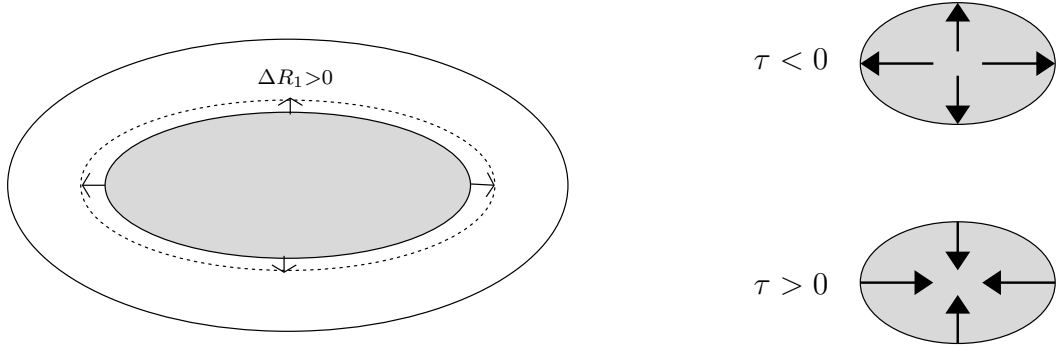}}
\caption{\small Deform the ring away from equilibrium by increasing its
(inner) radius, $\Delta R_1>0$. The backreaction creates a
conical-defect membrane (gray disk) inside the ring. A stable ring tends
to return to its original equilibrium radius, hence membrane {\it
pressure} ($\tau<0$) is needed to keep it static at
$R_1^\mathrm{equil}+\Delta R_1$. Instead, membrane {\it tension}
($\tau>0$) is required to halt the runaway increase of $R_1$ away from
equilibrium: an unstable configuration.}
\label{fig:radial}
\end{figure}

We regard the tension as a function of the $S^1$ radius, $\tau(R_1)$.
Given a black ring, we take it out of equilibrium by having
$R_1=R_1^\mathrm{equil}+\Delta R_1$ (fig.~\ref{fig:radial}).
There is obviously an ambiguity in the choice of $S^1$ radius $R_1$. We
will find, however, that for reasonable choices of $R_1$ the final results
are robust. For visualization purposes it is convenient to plot the
radial potential\footnote{Since $\tau$ actually gives the force acting
at each point of the ring, one might define the potential as $-\int 2\pi
R_1 \tau(R_1) \, dR_1$. The qualitative properties of these two
potentials are, however, the same.}
\beq\label{vofr}
V(R_1)=-\int^{R_1} \tau(R'_1) \, dR'_1\,,
\eeq
although we are only interested in its value around the
equilibrium points ($V'=\tau=0$).

Perturbing away from equilibrium, if we find
\beq\label{rinstab}
V'' = -\frac{d\tau}{dR_1}\Big|_\mathrm{equil}>0\,,
\eeq
(negative $\tau$ is pressure) then we infer that the equilibrium is
stable: in order
to keep the ring static at a larger radius $R_1^\mathrm{equil}+dR_1$, an
outward-pushing pressure has to be applied, indicating that the ring wants to
return to its original position. 

Conversely,
\beq\label{rstab}
V'' = -\frac{d\tau}{dR_1}\Big|_\mathrm{equil}<0\,
\eeq
is interpreted as the appearance of an inward-pulling tension needed to prevent
the runaway increase of the ring radius away from equilibrium, \ie
instability.

Before proceeding to the more general analysis, we study the instructive
regime of very thin rings, $\lambda,\nu\ll 1$, since in this limit we
recover a Newtonian result for the equilibrium condition.

\subsection{Mechanics of thin black rings}
\label{sec:balance}
When
$\lambda,\nu$ are very small we can approximate
\beq
M\simeq
\frac{3\pi}{4G}R^2\lambda\,,\quad \Omega\simeq
\frac{1}{R}\sqrt{1-\frac{\nu}{\lambda}}\,,\quad \tau\simeq\frac{3}{8
G}(\lambda-2\nu)\,,\quad R_1\simeq R\sqrt{\frac{\lambda}{\nu}}\,.
\eeq
In this limit there is no ambiguity in the definition of $R_1$. Eliminating
$\lambda,\nu, R$ from
these four equations, thus leaving only physical magnitudes, we find
\beq\label{tequil}
2\pi R_1\tau= -\frac{M}{R_1}+M\Omega^2
R_1\,.
\eeq
This equation has a neat interpretation. First
observe that since $\tau$ is the force that acts at each point along the
circle of the ring, the total force acting on it is
\beq
F=2\pi
R_1\tau\,.
\eeq
Next, $M/2\pi R_1$ is the energy per unit length of the
ring, so it is natural to define a string tension
\beq
\mathcal{T}=\frac{M}{2\pi R_1}\,.
\eeq
(As a cautionary note, recall that tension equals energy per unit length only when
there is boost-invariance along the string, which is not the case here).
Finally, the last term $M\Omega^2
R_1$ is obviously a centrifugal force. The equation \reef{tequil}
\beq\label{fequil}
F=-2\pi\mathcal{T}+M\Omega^2 R_1\,
\eeq
is then {\em exactly} the Newtonian result for the  equilibrium of a
rotating circular string of tension $\mathcal{T}$, total mass $M$,
radius $R_1$ and angular velocity $\Omega$, under an external
centripetal
force $F$.\footnote{The normalization of $\tau$ in \reef{taudelta} was
chosen so that in the absence of string tension we recover the correct
force law $F=M\Omega^2 R_1$.}

The condition for equilibrium in the absence of the external force
fixes
\beq\label{equilt}
\mathcal{T}=\frac{M\Omega^2 R_1}{2\pi}\, .
\eeq
Hence at equilibrium,
\beq\label{equilom}
\Omega R_1=1\,,
\eeq
which was observed early on \cite{ER}.

Notice that the gravitational self-attraction of the ring is absent in
\reef{fequil}. This force is proportional to $GM^2$ and therefore
appears only if we expand to $O(\lambda^2)$. So gravitational forces are
only a subleading effect in the balance of very thin black rings. An important
consequence of this is the plausibility of the existence of black rings
in higher dimensions\footnote{Ref.~\cite{HovMyers} also reached this
conclusion from a similar argument but the details of their mechanical
model are different.}. The tension and centrifugal forces, being
confined to the plane of rotation where the ring lies, are independent
of the total number of dimensions --- unlike the gravitational force,
which is strongly dimension-dependent but negligible for very thin black
rings.
So we expect the existence of equilibrium configurations of thin
rotating black rings in any dimension $D>5$.\footnote{Note that: (i) the argument
does not apply to $D=4$, since there are no black strings to begin with;
(ii) in $D>5$, taking a ring out of equilibrium is expected to create a stronger
singularity than a conical defect.}

A simple argument, based on the phase diagram of fig.~\ref{fig:phase},
suggests that thin black rings should be radially stable\footnote{We
thank Veronika Hubeny for suggesting this.}. Consider, for fixed total
mass and angular momentum, a slight increase in the radius $R_1$ away
from the equilibrium value. For stationary equilibrium solutions, $j$
increases as the ring radius grows larger, so a ring at larger radius
and the same mass requires larger angular momentum to balance itself.
Therefore, the radially perturbed solution has less angular momentum
than required for balancing the tension, and it would tend to be pulled
back to equilibrium. I.e., the thin black ring should be radially
stable. This is also the expectation from the Newtonian balance of
forces.

\subsection{Radial stability} \label{sec:radial-general}

We now apply the method of off-shell radial perturbations to study the
stability of black rings, thin as well as fat.

Let us discuss first the practical implementation.
First of all, we have to choose a measure for the circle radius.
It seems appropriate to consider the inner ring radius $R_1^\mathrm{inner}=
R_1^{x=1}$, since we intend to study whether the inner hole of the ring
tends to open up or close in as we vary its radius. So we will choose
(see \reef{r1x})
\beq
R_1=R^\mathrm{inner}_1=R\sqrt{\frac{\lambda}{\nu}}\,.
\eeq
As it turns out, qualitatively the results do not depend on the choice of $R_1$
as long as this radius behaves monotonically as we move from the thin
to the fat ring branch (as is the case for $R^\mathrm{inner}_1$
and also for $R_1^\circ$).\footnote{Other choices of $R_1$ which
diverge in the
nakedly singular limit $\nu \to 1$, such as the outer radius at
$x=-1$, give bizarre-looking  
results for fat rings in this
limit. These are most probably simply
artifacts of a bad choice of $R_1$, but one should keep in mind that
they might signal a lesser reliability of our method in the $ \nu \to
1$ limit.}

We must also specify which quantities we should keep fixed while we take
the ring out of equilibrium. These should be quantities that we regard
as intrinsic to the black ring as a physical object. It seems natural to keep
$J$ fixed: recall that, in a microscopic interpretation with quantized
parameters, $J$ measures the number of units of momentum along the
direction of the ring. One might also want to keep $M$ fixed. This is
slightly tricky since the ADM mass $M$ measured at infinity receives a
contribution from the conical disk membrane that is present away from
equilibrium. A different possibility is to keep the horizon area fixed.
This sounds sensible, since the area, \ie entropy, is a measure of the
number of ring microstates. Keeping it fixed amounts to an adiabatic
change.

The adiabatic variation of the radius seems to be the most
sensible choice, however, it is just as easy to study both
possibilities, \ie keeping fixed either area and spin, or mass and spin, and
so we consider both in the following. Happily, our conclusions turn out
not to depend on the choice.

If we vary the solution parameters $R$, $\lambda$ and $\nu$ in such a way
that $J$ and ${\cal A}_H$, or $J$ and $M$, remain fixed, then there is
only one independent variation parameter, say $\nu$. It is again
convenient to work with reduced quantities, and to fix the scale
of the black rings by fixing one physical magnitude. Since in any case we want
to keep $J$ fixed, we introduce a reduced radius
\beq
r=\frac{R_1}{J^{1/3}}\,.
\eeq

If we want to keep $J$ and $M$ fixed then the
parameter to use is the reduced spin $j$ introduced in
\reef{jandaH}. Keeping it fixed, $dj=0$, yields
\beq\label{jfixed}
\left(\frac{d\lambda}{d\nu}\right)_{M,J}=
-\frac{\partial j/\partial\nu}{\partial j/\partial\lambda}=
\frac{1-\nu^2}{\nu(1+\nu^2)^2}\,.
\eeq
Since we are interested in perturbations around equilibrium
configurations, we evaluate the derivatives at equilibrium,
$\lambda=\lambda_c$ as given in \reef{balance}.

For the purpose of keeping $J$ and ${\cal A}_H$ fixed we introduce
a second reduced area,
\beq\label{ahat}
\hat a_H=\frac{{\cal A}_H}{J}\,,
\eeq
so that
\beq\label{afixed}
\left(\frac{d\lambda}{d\nu}\right)_{{\cal A}_H,J}=
-\frac{\partial \hat a_H/\partial\nu}{\partial \hat a_H/\partial\lambda}=
\frac{2(2-\nu)(1-\nu)}{(1+\nu^2)^2}\,,
\eeq
again evaluated at equilibrium.

We can now use these results to compute
\beq\label{dtaudr}
\left(\frac{d\tau}
{dr}\right)_{\ast,J}=
\frac{\left(\frac{d\tau}{d\nu}\right)_{\ast,J}}{\left(\frac{dr}{d\nu}\right)_{\ast,J}}
=\frac
{\partial_\nu\tau+\left(\frac{d\lambda}{d\nu}\right)_{\ast,J}\partial_\lambda\tau}
{\partial_\nu r+\left(\frac{d\lambda}{d\nu}\right)_{\ast,J}\partial_\lambda r}\,,
\eeq
(with $\ast$ being $M$ or ${\cal A}_H$) whose sign will give us,
according to the above discussion, the radial stability properties of
the black ring.

To see what the sign is, first note that the numerator in
\reef{dtaudr},
\beq
\left(\frac{d\tau}{d\nu}\right)_{M,J}=\frac{3}{8
G}\frac{1-2\nu}{\nu(1- \nu^2)}\,, \eeq or \beq
\left(\frac{d\tau}{d\nu}\right)_{{\cal A}_H,J}=\frac{3}{4
G}\frac{1-2\nu}{(1- \nu)(1+\nu)^2}\,,
\eeq
in both cases changes
sign at $\nu=1/2$ (and only there), which corresponds to the
minimally spinning ring at the cusp where the two branches meet.

Next, the denominator in \reef{dtaudr},
\beq\label{drdnu}
\left(\frac{dr}{d\nu}\right)_{M,J}=
-\frac{r}{2\nu(1-\nu)}\,,
\eeq
or
\beq\label{drdnu2}
\left(\frac{dr}{d\nu}\right)_{{\cal A}_H,J}=
-r\frac{1+5\nu-\nu^2+\nu^3}{3\nu(1-\nu^2)(1+\nu)}
\,,
\eeq
is in both cases negative over the entire
parameter range $0\leq \nu< 1$.

So for this family of radial perturbations we conclude that
\begin{itemize}

\item $\displaystyle\left(-\frac{d\tau}{d
r}\right)_{\ast,J}>0$ for $0\leq\nu<1/2$: {\it thin black rings are radially stable.}

\item $\displaystyle\left(-\frac{d\tau}{d
r}\right)_{\ast,J}<0$ for $1/2<\nu<1$: {\it fat black rings are
radially unstable.}

\end{itemize}
The conclusion is independent of whether we fix $\ast=M$ or $\ast={\cal
A}_H$.

\begin{figure}[t!]
\begin{picture}(0,0)(0,0)
\put(29,50){$\scriptstyle{V}$} 
\put(140,61){$\scriptstyle{r}$}
\put(83.6,63){$\scriptstyle{1}$} 
\put(131.1,63){$\scriptstyle{2}$}
\put(33.8,60.1){$\scriptstyle{0}$}
\put(30,30.5){$\scriptstyle{-0.1}$}
\put(30,0.9){$\scriptstyle{-0.2}$}
\put(144,28){$\scriptscriptstyle{27/32}$}
\put(144,30){$\scriptscriptstyle{0.9}$}
\put(144,32){$\scriptscriptstyle{0.95}$}
\put(144,34){$\scriptscriptstyle{1.}$}
\put(144,36){$\scriptscriptstyle{1.05}$}
\put(144,38){$\scriptscriptstyle{1.1}$}
\put(139,40){$\scriptscriptstyle{j^2=1.2}$}
\put(92.5,16){\vector(-1,4){1.6}} \put(86,13){\scriptsize{thin ring}}
\put(70.5,50){\vector(-3,-4){13}} \put(71.5,50){\scriptsize{fat ring}}
\end{picture}
\centerline{
\includegraphics[width=10cm]{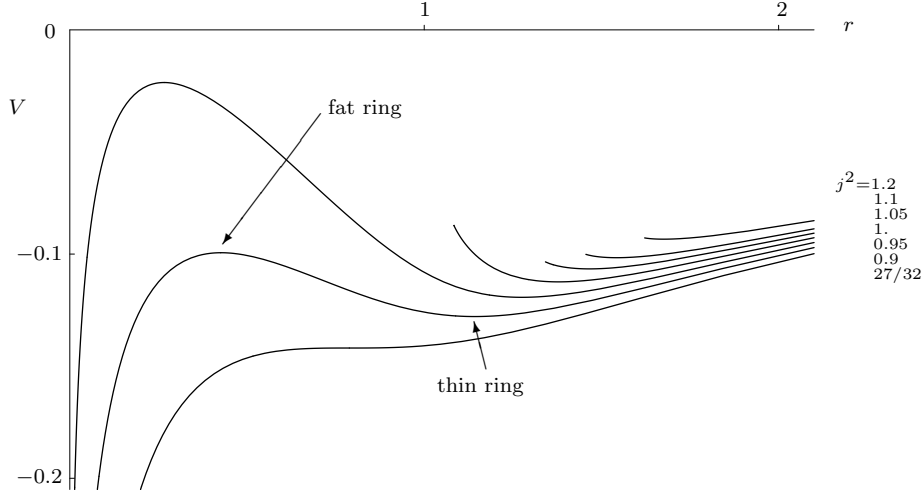}}
\caption{\small Radial potential $V(r)$, for fixed values of the
mass and the spin, \ie for fixed $j$. Fat black rings correspond to
unstable equilibrium at local maxima of the potential,
and thin black rings to stable local minima. Observe
that, in agreement with the phase diagram of fig.~\ref{fig:phase},
the two branches exist when $\sqrt{27/32}<j<1$, and they merge at
$j=\sqrt{27/32}$. When $j^2>1$ there is a minimum value of the
radius for given $M,J$, hence the abrupt termination of the
corresponding curves.} \label{fig:VofR}
\end{figure}

Figure \ref{fig:VofR} shows radial potentials for representative
values of constant $j$. When $\sqrt{27/32}<j<1$, the potential
has a local minimum corresponding to the (radially) stable thin black
ring, and a local maximum at a smaller value of $r$ corresponding to the
unstable fat black ring. When $j = \sqrt{27/32}$ the potential
has an inflection point, so the black ring with minimal spin $j$ will
also be unstable to radial perturbations. As $j \to 1$, the local
maximum of the fat black ring goes to $\infty$, and disappears at
$j=1$. Thus for $j \ge 1$, only the local minimum exists corresponding
to a radially stable thin black ring with large angular momentum.

The potentials for fixed spin and area are qualitatively similar, with
fat black rings sitting at maxima of the potential higher than the
minima for thin black rings. Note that the difference in the potential is
consistent with the fact that, for fixed spin and area, thin black rings
have smaller mass than fat ones.

The shape of the potential suggests that fat black rings can either
collapse into a MP black hole, or expand into a thin black ring. If the
mass and spin are conserved during the evolution, both possibilities are
allowed by the area law. Whether the corresponding MP black hole and
thin black ring can themselves be the stable endpoint of the evolution
under a generic perturbation is an open issue. A possibility suggested
by the phase diagram, fig.~\ref{fig:phase}, is that the final state is
the solution that maximizes entropy globally. However, the classical
dynamical evolution needs not conform to this expectation. Thin black
rings very close to the cusp may be metastable configurations.

As for the minimally spinning ring, the only plausible evolution, also
suggested by the shape of the potential, is the collapse into a MP black
hole. This is also consistent with the instability found in
\cite{grf}, which we will rederive in sec.~\ref{sec:geodesics} below.

The existence of a classical instability for fat black rings, and the
change of stability exactly at the cusp at $j=\sqrt{27/32}$, is in
agreement with the results of \cite{loar}. We discuss this in more
detail next.


\subsection{Comparison to the `turning-point' method}
\label{sec:turning}
                                                                                
We have found that the radial stability of the black ring changes at
the minimally spinning ring at the cusp at $\nu=1/2$. On this
configuration, the radial perturbation can be said to have a zero-mode
at linearized level. This feature could in fact have been anticipated
without any calculations. To see this, let us discuss the
perturbations with $M$ and $J$ fixed --- the case with ${\cal A}_H$ and
$J$ fixed will easily be seen to share the same features. 
                                                                                
The configurations of black rings that we are studying correspond to
points in the plane $(\nu,\lambda)$, with $\nu,\lambda\in [0,1]$.
Equilibrium configurations lie on the curve $\lambda=\lambda_c(\nu)$
in eq.~\reef{balance}. In our analysis above, we move off-shell away
from this curve to a point $(\nu+d\nu,\lambda_c+d\lambda)$ in such a
way that $j$ remains fixed, $dj=0$. This determines the directional
derivative $(d\lambda/d\nu)_{M,J}$, eq.~\reef{jfixed}, along which we
move away from the equilibrium curve. 

Now consider instead that we move along the on-shell, equilibrium
curve. Then $j$ reaches a minimum at the cusp, so
$(dj)_\mathrm{equil}=0$ at $\nu=1/2$. This means that, at the cusp,
the radial deformation with fixed $j$ does not take the ring off-shell
but instead moves it to another, infinitesimally nearby, equilibrium
solution. So at $\nu=1/2$, $(d\lambda/d\nu)_{M,J}=d\lambda_c/d\nu$,
\ie we move tangentially to the equilibrium curve and not away from
it, at least at linear order. Hence, there is a linearized marginal
(zero-mode) deformation at the cusp. The same argument applies to
variations with fixed ${\cal A}_H$ and $J$, since $\hat a_H$ in
\reef{ahat} also has an extremum at $\nu=1/2$. 

This kind of reasoning is closely related to a previous analysis of 
black ring stability by Arcioni and Lozano-Tellechea in
ref.~\cite{loar}. It is illuminating to discuss in more detail the
connections, and differences, between their approach and ours.

The analysis of ref.~\cite{loar} uses rather generic arguments based on
the `Poincar\'e (or `turning-point') method'. This method only requires
qualitative knowledge of the equilibrium curve in the phase diagram.
This curve corresponds to extrema of some thermodynamic function, say
the entropy in the microcanonical ensemble. By consideration of the
possibility of perturbations away from equilibrium, one can conclude
that the presence of `turning-points' along the equilibrium curve implies
the existence of (at least) one zero mode for {\it some kind} of
perturbation. The argument is, in fact, exactly the same as we have just
given above, only with the difference that we have provided an explicit
kind of perturbations away from equilibrium, namely, those that occur
when we vary $\lambda$ and $\nu$ independently. The analysis of
\cite{loar} is more general than ours in that it does not need to
specify the form of the perturbation (note that the argument we have
given above does not actually use this information either, but is based
exclusively on the fact that $j$ reaches a minimum along the equilibrium
curve), but then it is more abstract since it does not allow to identify
the nature of the instability. In our approach, we have not only
specified a particular kind of off-equilibrium deformations, but also
identified them as radial deformations. This is made particularly
precise from \reef{drdnu}, which implies that increasing $\nu$ with
fixed $M$ and $J$ always results in decreasing $R_1$. Hence we can claim
that we are perturbing the system by varying the (inner) radius of the
ring.

Another difference between our approach and that of \cite{loar} concerns
the way in which the direction of the change from stability to
instability at the turning point is determined. Our arguments can be
rephrased as follows. On the plane $(\lambda,\nu)$, on one side of the
equilibrium curve $\tau$ is negative, and positive on the other side. So
as we deform the ring away from equilibrium, the sign of $d\tau$ will
depend on which of the two regions of the plane we are moving into. This
direction is given by $(d\lambda/d\nu)_{M,J}$, which can be written as
\beq
  \left(\frac{d\lambda}{d\nu}\right)_{M,J} 
  = \frac{1}{2\nu}\; \lambda_c'(\nu) \, .
\eeq
We see that for $0<\nu<1/2$ this directional derivative is steeper than
the slope of the tangent of the equilibrium curve
$\lambda=\lambda_c(\nu)$, and therefore the directional derivative takes
us to a point above the curve $\lambda_c$, where $\tau$ is negative. For
$1/2 < \nu < 1$, the tangent of the equilibrium curve is steeper than
the direction of the perturbation and therefore the latter takes us to
the $\tau>0$ region below the equilibrium curve. Thus stability {\it
does change} exactly at $\nu=1/2$ (and only there), where it is, to linear order,
indifferent. To determine now which side of the curve is stable and
which unstable, it only remains to correlate these directional derivatives
to radial variations, using \reef{drdnu}, and interpret $d\tau/dr$ as we
have done above. Note that our analysis does require explicit knowledge
of derivatives of $V(r)$ in directions away from the equilibrium curve,
and hence of properties of the off-shell configurations. 

The reasoning in \cite{loar}, however, cannot use this off-shell
information since the nature of the perturbation away from equilibrium
remains unspecified. Instead, they (essentially) use the fact that if we
move away from the turning point into the thin black ring branch, the
entropy is larger than if we move into the fat branch. By continuity,
the former corresponds to local maxima of the entropy (hence stable) and
the latter to local minima (hence unstable). 

Crucially, observe that the criteria for stability are actually
different in the two approaches. In ref.~\cite{loar} equilibrium
configurations are regarded as maxima or minima of a microcanonical
potential, \ie the entropy. Our criterion, instead, determines
equilibrium and stability using the (off-shell) radial potential $V(r)$.
It is satisfying that these two approaches, one thermodynamical, the
other mechanical, consistently yield the same results.

In conclusion, our analysis of radial stability confirms and gives a
different, more mechanical perspective on the approach of \cite{loar}.
Furthermore, it identifies the nature of the instability of the fat ring
branch as due to radial perturbations.



\setcounter{equation}{0}
\section{Gregory-Laflamme Instability and Fragmentation}
\label{sec:gl}

\subsection{Gregory-Laflamme instability in black rings}

A static compactified black string $S^2 \times S^1$
suffers from a long-wavelength Gregory-Laflamme instability
\cite{GL1,GL2} if the
size the two-sphere is small compared with the size of the circle.
Introducing the ratio of the radii $k = R_2/R_1$, the critical value
is $k_\rom{GL} \approx 0.88$, so that for $k \le
k_\rom{GL}$ the black string is unstable.

We compare the sizes of the horizon $S^2$ and $S^1$ for the black rings.
For large angular momenta $j$ (small $\nu$) the ratio of the $S^2$ and
$S^1$ radii is small, since $R_2 \sim \nu R$ while
$R_1 \sim \sqrt{2}R$. Hence we expect that black rings are
unstable, at least in the thin ring regime, to a Gregory-Laflamme type
perturbation. With the various radii defined in the previous section, we
can estimate $k$ as a function of the `shape parameter' $\nu$. However,
we stress that this will only give a rough estimate. The distortion of
the horizon makes the black ring different from the black string and we
cannot expect to trust the results for general $\nu$. Based on the
results of section \ref{sec:shape}, we would certainly not trust an
estimate for fitting a black string Gregory-Laflamme mode on a black
ring for $\nu$ beyond $\nu \sim 0.5$, \ie we do not trust the estimates
for fat black rings.

Since the black ring has angular momentum, we must correct for the
effect of the velocity. Hovdebo and Myers \cite{HovMyers} have shown
that for a boosted black string the threshold wavelength can be
computed from the static mode by simply including a relativistic
contraction; the result is \be
  k_\rom{GL}(v) = \frac{1}{\sqrt{1-v^2}} k_\rom{GL}(0) \, ,
\ee where $v=|\tanh \beta|$ is the velocity of the black string with
boost parameter $\beta$. The boost increases the threshold, making
`more' black strings unstable.

\begin{figure}
\begin{picture}(0,0)(0,0)
\put(77,57){$\scriptstyle{k_\rom{BR}}$}
\put(135,-1){$\scriptstyle{j}$}
\put(39,-4){$\scriptscriptstyle{\sqrt{\frac{27}{32}}}$}
\put(79,-3){$\scriptscriptstyle{1}$}
\put(121.6,-3){$\scriptscriptstyle{1.1}$}
\put(81.5,27){$\scriptscriptstyle{0.5}$}
\put(82.5,51){$\scriptscriptstyle{k_\rom{GL}=0.88}$}
\put(40,51){{\scriptsize GL-stable}}
\put(40,45){{\scriptsize GL-unstable}}
\end{picture}
\centerline{
\includegraphics[width=9cm]{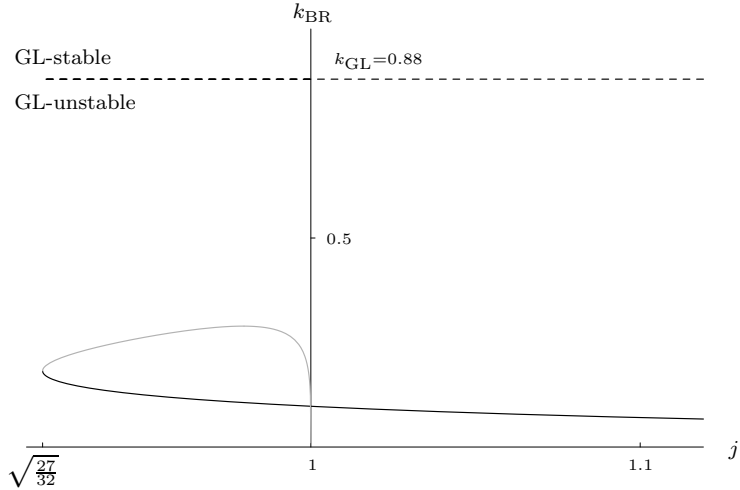}}
\smallskip
\caption{\small The ratio $k_\rom{BR}=\sqrt{1-v^2} R_2/R_1$ versus
$j$ for black rings (solid curve) is always well below the
Gregory-Laflamme threshold $k_\rom{GL}\simeq 0.88$ (dashed line), so
we expect black rings to be unstable to GL-type modes. Here we use
$R_1^\circ$ and $R_2^\circ$ for the radius of the $S^1$ and $S^2$. The
plot includes the results for the fat black ring branch (gray), although
we do not trust the results in this range.
}
\label{fig:GLfit}
\end{figure}

Taking the velocity correction into account, we compare for black
rings the ratio of $R_2/R_1$ at given $j$ to the appropriate threshold
$k_\rom{GL}(v)$. We choose the radii measured at the `equator' of the
$S^2$ (eq.~\reef{fatxo}), and we use eq.~\reef{velocity} to calculate
the velocity $v$ (the results for thin black rings are very similar
for all other choices of $R_1$, $R_2$). The practical implementation
is most easily done by comparing
\be
  \label{ratio}
k_\rom{BR}\equiv  \frac{R_2}{R_1}\sqrt{1-v^2}
\ee
with $\sqrt{1-v^2}\,k_\rom{GL}(v) \approx  0.88$.
Figure \ref{fig:GLfit} shows \reef{ratio} plotted versus $j$.
Thin black rings have $k_\rom{BR}$ smaller by a factor of
three or more than the GL threshold $k_\rom{GL}(0)\simeq 0.88$, and
hence they are expected to be unstable.

Strikingly, there is always a very wide margin to accommodate
unstable modes. This, together with our analysis of the shape of
black rings (fig.~\ref{fig:S2dist}), strongly suggests that the
instability could extend down to values of $j$ very close to its
minimum, hence covering most, if not all, of the branch of thin
black rings.

\subsection{Fragmentation into black holes}
\label{sec:frag}

If the thin black rings are unstable to Gregory-Laflamme
instabilities, then what is the endpoint of the decay? The answer
hinges on the unresolved issue of the fate of the black string
instability \cite{HM, luisetal, DMnew,revbhbs}. For black rings,
there is the additional effect that the growing inhomogeneities will
rotate with the ring, causing a varying quadrupole moment and hence
gravitational radiation. However, the timescales for the evolution
of the GL instability of thin strings are very short: the inverse
frequency of a typical GL unstable mode is of the order of
\beq
\label{tgl} T_{GL}\sim \frac{GM}{R}\,
\eeq
(which seems to
remain valid at least insofar as the non-linear evolution has been
followed \cite{luisetal}). On the other hand, gravitational
radiation is a notoriously inefficient process. We can estimate the
rate at which the black ring will lose energy through gravitational
radiation by using a rough adaptation to five dimensions of the
results from the quadrupole formula in \cite{MTW} (sec.~36.2) . The
timescale for a system to radiate a fraction of its mass is
\beq
T_{gw}\sim \Omega^{-1} L_{internal}^{-1}
\eeq
where the black ring
rotation velocity is $\Omega\sim R^{-1}$ and the dimensionless
$L_{internal}$ characterizes the radiating power
and can be estimated to be $\sim GM/R^2$.
Hence\footnote{Compared to the systems analyzed in \eg sec.~36.4 of
\cite{MTW}, in our case there is no virialization between kinetic
and gravitational potential energy, since, as we have seen, the
latter is negligible. This makes the power bigger (since there is a
force stronger than gravitation binding the system), but still the
result is very small.}
\beq
\label{tgw}
T_{gw}\sim \frac{R^3}{GM}\,.
\eeq
Since $\sqrt{GM}\ll R$ for thin rings,
\beq
\frac{T_{GL}}{T_{gw}}\sim \frac{(GM)^2}{R^4}\ll 1\,,
\eeq
so the
effects of gravitational radiation take too long to become
noticeable and hence are negligible --- unless the black string
enters a regime where the evolution of the instability dramatically
slows down. Although some slowing-down is observed in the non-linear regime
\cite{luisetal}, there is no evidence that it will prevent the simplest
possibility, \ie that the black string continues to pinch down to
Planck-scale necks in a finite time of order $T_{GL}$ (measured by late
asymptotic observers), at which point, through quantum gravity effects,
it fragments into black holes of spherical topology.

For the black ring, this implies that it will fragment into a number of
black holes flying apart. The simplest way to relate this fragmentation
to the GL instability follows the observation in \cite{GL1, GL2} that
the unstable regime roughly coincides with the range of parameters where
the entropy of split black holes is higher than that of the original
black string. A similar argument was used in \cite{EM} to argue that
higher-dimensional ultraspinning black holes may be unstable. In this
section, proceeding along the same lines, we argue that fragmentation of
the black ring is entropically possible and so indicates that a GL-like
instability may exist for black rings.

The set up for the calculation is actually the same whether the initial black
hole is a black ring or a Myers-Perry black hole. We consider as the
initial state a black hole with mass $M$ and angular momentum $J$. The
final state consists of $n$ black holes, each of mass $m$, infinitely
far away from each other and moving in various directions, so that the
total momentum is zero and the total angular momentum is $J$. We take
the final black holes to be non-spinning, since this maximizes the area
of the black holes.

Suppose the splitting occurs with an impact parameter ${\cal R}$. Then,
in the center of mass frame,
the magnitude of momentum of each of the final black holes is $J/(n {\cal
R})$. Assuming, as is usual in these crude calculations, that no energy is lost
to radiation in the process of the split-up, we have
\begin{equation}
M = n \sqrt{m^2 + \frac{J^2}{n^2 {\cal R}^2}}\, .
\end{equation}
Solving for $m$ gives
$$
m = \frac{1}{n}\sqrt{M^2 - \frac{J^2}{{\cal R}^2}} \, .
$$
The total area of the final state black holes is
\be
  \mathcal{A}_\rom{split}=  2\pi^2\, n \,
  \left(
    \frac{8G}{3\pi} \, m
  \right)^{3/2} \, ,
\ee and so in terms of the dimensionless horizon area we find \be
 \label{asplit}
  a_\rom{split} = \frac{3}{16} \sqrt{\frac{3}{\pi}} \frac{
\mathcal{A}_\rom{split}}{(GM)^{3/2}}
 = \frac{2\sqrt{2}}{\sqrt{n}}
 \left(
   1- \frac{J^2}{M^2 \mathcal{R}^2}
 \right)^{3/4} \, .
\ee

The fragmentation of a $D$-dimensional highly spinning Myers-Perry black
hole as the initial state was considered in \cite{EM}. For $D=5$, if for
simplicity\footnote{This is actually a conservative value. Taking the
size of the horizon in the plane of rotation would give a larger value
for $\mathcal{R}$ \cite{EM}.}, we take $\mathcal{R}$ to be equal to the
rotation parameter $a=3J/2M$, then the configuration of $n$ fragmented
black holes dominates entropically for $j^2 > 1- 5\sqrt{5}/(27 n)$. For
example, for $n=2$, the split two black hole system has $a_\rom{split}
\approx 1.29$ and this is higher than the Myers-Perry black hole entropy
when $j > 0.89$ (compare with $a_\rom{H}^\rom{MP}$ vs.~$j$ in figure
\ref{fig:phase}). Note that all black rings have $a_\rom{H} \le 1$.
Although our approximations are rather crude and these numerical
values are therefore not too reliable, the estimates suggest that
for $j$ close to its maximum value, the Myers-Perry black hole becomes
unstable, and rather than decaying to a black ring, it could fragment
into two non-spinning black holes flying apart.

We now take as the initial state a balanced black ring parameterized by
$\nu$ and $R$. It is natural to take the splitting radius (impact
parameter) $\mathcal{R}$ to be close to the radius of the $S^1$ of the
black ring. As discussed in section \ref{sec:shape} and appendix
\ref{app:radii}, this radius depends
on $x$, and using $\mathcal{R}=R_1^x$  (eq.~\reef{r1x}) we
find that the total horizon area \reef{asplit} of the final state of $n$
black
holes is
\begin{equation}
  \label{aHsplit}
  a_\rom{split}=\frac{2 \sqrt{2}}{3 \sqrt{3n }} \left(\frac{8 - \nu
  \left(\nu ^2+\nu +2 x (\nu +1)+10\right)}{1- \nu }\right)^{3/4}\,.
\end{equation}
Presumably, we can only apply this approach for thin black rings
($0<\nu<1/2$), where the horizon distortion is small and the results do
not depend much on the choice of $x$ for the radius $R_1^x$.

Comparing $a_\rom{split}$ with the reduced area of the black ring
$a_\rom{H}^\rom{BR}$ of \reef{jandaH2}, we find that
fragmentation is always entropically possible.
This is in agreement with
what we have found from fitting GL modes inside a thin ring. The
agreement between both models is in fact fairly good, since there is a
nice correlation between the estimate that each of the models gives for
the maximum number $n$ of black holes that a given black ring can split
into. Taking the splitting radius to be $\mathcal{R}=R_1^\circ$, the
maximum number $n_\rom{split}$ of black holes which a black ring can
fragment into while increasing the entropy in the process is
\be n_\rom{split} =
\left\lfloor \left(\frac{a_\rom{split}(x = x_\circ, n =
1)}{a_\rom{H}^\rom{ring}}\right)^2 \right\rfloor \, ,
\ee
where $a_\rom{H}^\rom{ring}$ is computed using \reef{AH} and
\reef{jandaH}. This can be compared to the maximum number of
Gregory-Laflamme threshold modes $n_\rom{GL}$ that can be fit on the
black ring
\be \label{numGLmodes}
  n_\rom{GL} =
  \left\lfloor
  \frac{R^\circ_{1}}{R^\circ_{2}} \frac{ 0.88}{\sqrt{1 -
v^2}}\right\rfloor \,.
\ee

\begin{figure}[t]
\begin{picture}(0,0)(0,0)
\put(36,63){$\scriptstyle{n}$}
\put(140,-1){$\scriptstyle{j}$}
\put(34,11.7){$\scriptscriptstyle{5}$}
\put(33,24.4){$\scriptscriptstyle{10}$}
\put(33,36.6){$\scriptscriptstyle{15}$}
\put(33,48.8){$\scriptscriptstyle{20}$}
\put(35,-4){$\scriptscriptstyle{\sqrt{\frac{27}{32}}}$}
\put(52.7,-3){$\scriptscriptstyle{1}$}
\put(134.5,-3){$\scriptscriptstyle{1.5}$}
\put(116,46){$\scriptstyle{n_\rom{split}}$}
\put(104,56){$\scriptstyle{n_\rom{GL}}$}
\end{picture}
\centerline{
\includegraphics[width=10cm]{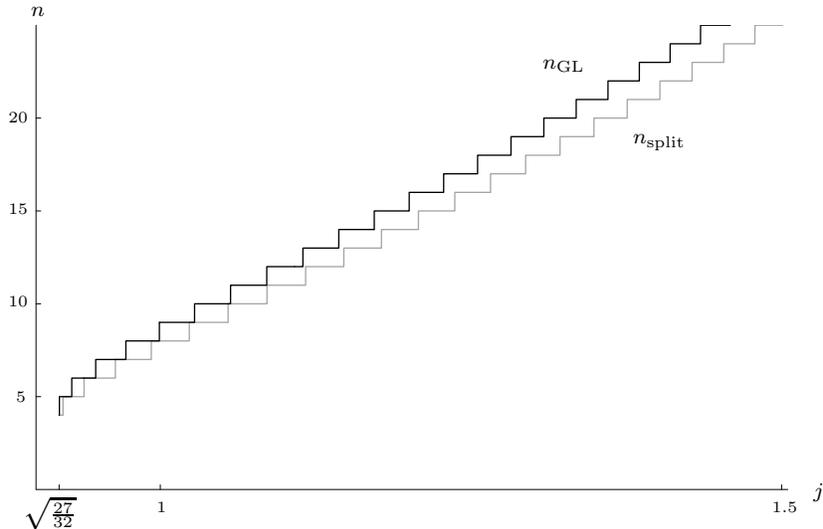}}
\caption{\small We see a strong correlation in the maximum number of
black holes that a black ring can split into, as a function of $j$, estimated in the two
models described in the text: $n_\rom{split}$ in the fragmentation model (gray) and
$n_\rom{GL}$ (black) in the GL-instability model.
The black ring radii are always estimated as $R_{1}^\circ$ and $R_{2}^\circ$.}
\label{fig:GLsplit}
\end{figure}

Figure \ref{fig:GLsplit} exhibits $n_\rom{GL}$ and $n_\rom{split}$ as
functions of $j$ for thin black rings ($0<\nu<1/2$), and shows that they
are fairly well correlated. The correlation is in fact not wholly unexpected,
given that the fragmentation model for black strings already gives a
good estimate for the wavelength of the threshold GL mode, and that, as
we have seen, thin black rings are well approximated as black strings of
finite length.

It must be noted, though, that neither $n_\rom{GL}$ nor $n_\rom{split}$
are the expected number of black holes that the black ring splits into,
even within the crudeness of the models involved. Physically, it makes
sense to assume that the pinching-off of the black ring will be
dominated by the fastest-growing GL modes, \ie the modes with the
largest imaginary frequency. The wavelength of these modes is slightly
more than twice the wavelength of the threshold (zero-frequency) mode,
so typically this will result in roughly half a number of pinches and
therefore half the number of final black holes than $n_\rom{GL}$ above.
In fact, the entropically-favored number of final black holes is not as
many of them but instead as few as possible! This minimum number is
obviously two.\footnote{Or, perhaps, one ejected black hole plus
radiation in the opposite direction.} However, typically this requires a
GL mode of longer wavelength than the fastest mode. So even if
fragmentation into just two black holes may be entropically favored, it
will be dynamically suppressed.

So, presumably, the black ring will fragment into a number of black
holes dominated roughly by the fastest unstable mode, and this fragmentation will
{\it not} maximize the final entropy. The split black holes will fly
apart and probably will not have the chance to merge into only two black
holes to maximize the entropy.


\setcounter{equation}{0}
\section{Emission/Absorption of Massless Particles}
\label{sec:geodesics}

An oft-studied way to test the dynamical stability of black holes is via
absorption or emission of small-energy particles \cite{wald}. A neutral
particle carries away (or dumps in) an energy $\delta E$ and angular
momentum $\delta J$, which we assume are small enough to treat it as a
test particle moving on a geodesic. An initially stationary black
hole is then expected to evolve into an infinitessimally nearby stationary
solution with mass $M+ \epsilon\, \delta E$ and angular momentum
$J+\epsilon\, \delta J$
($\epsilon=+1$ corresponds to absorption and
$\epsilon=-1$ to emission of a particle with energy $\delta E>0$). This
evolution is classically possible only if the area of the event horizon
does not decrease. Then the first
law,
\be
  \label{deltaS}
\frac{\kappa}{8\pi G}\,\delta \mathcal{A}_H = \epsilon (\delta E - \Omega\, \delta J)
\ee
allows to discriminate between processes that are classically allowed or
forbidden.

Consider the absorption of a test particle. It may happen that no
infinitessimally nearby, stationary black hole solution is available as
a final state (\eg the solution with those final parameters could have
naked singularities), at least not without violating the area law. In
this case, the initial black hole cannot smoothly settle into a new
stationary state, and the absorption of the particle must be followed by
violent backreation --- \ie an instability.

It is often illuminating to also consider the reverse process in which
spontaneous emission of test particles takes place --- this is of course
only a very simplistic model for radiation. Imagine that it is
possible for a black hole to `emit' a test particle carrying away $\delta
E$ and $\delta J$ in such a way that the horizon area increases after
emission. This would be a strong suggestion that the initial black hole
is unstable, since it could increase its entropy through spontaneous
decay.

When the emitted/absorbed particle is neutral, it is
very easy to establish a criterion for these instabilities \cite{EM}.
On general
grounds, one expects a relation between the particle's energy and spin
of the form
\beq
\alpha\, \delta E = \Omega\, \delta J\,.
\eeq
The dimensionless `efficiency coefficient' $\alpha$ parameterizes the
efficiency at which the angular momentum is increased (or decreased) in
the process of absorbing/emitting a particle. Note that $\alpha$ is
positive (negative) for co-\mbox{(counter-)}rotating geodesics, so that
$\delta E>0$ in all cases.
Plugging this into \reef{deltaS},
\be
  \label{deltaS2}
\frac{\kappa}{8\pi G}\,\delta \mathcal{A}_H
  = \epsilon (1- \alpha) \delta E
\ee
we see that iff $\alpha\geq 1$, the area law permits spontaneous
emission ($\epsilon =-1$), and an instability is expected. Conversely,
the absorption ($\epsilon =+1$) of such a particle cannot happen without
violating the area law, again signalling the impossibility of smooth
classical evolution.

For the purposes of this paper, there are three different kinds of
situations where one is motivated to study emission and absorption of
test particles:

The first one is as a simple model for an instability of very fastly
rotating black objects, as in \cite{EM}. An arbitrarily large rotation
suggests that the object might spontaneously radiate away angular
momentum. This process may also be taken as a simple model for the
evolution of the Gregory-Laflamme instability of sec.~\ref{sec:gl}. We
studied above the possibility of fragmentation, but another possible
evolution is that the inhomogeneities along the ring cause radiation of
gravitational waves, which slow down the ring. We may hope to model
this crudely
as spontaneous emission of massless test particles. Observe that the
efficiency parameter $\alpha$ is maximized for a massless particle that
tangentially skims the horizon's edge in corotating sense, and
therefore, below we will focus on this case.

The second unstable regime we envisage is in a certain sense the
opposite, since we are now interested in {\it minimizing} the efficiency
parameter. Consider the black ring at the cusp between branches, with
$j=j_{min}$. Imagine dropping into it a particle with the effect that
$\delta j<0$. Since there are no black rings with $j<j_{min}$, we are in
one of the situations considered above, where the system must backreact
violently. We will show that there do exist null geodesics with the
required properties. Actually, there are many classes of such geodesics
(massive particles yield an even more negative $\delta j$),
but it will suffice for us to restrict ourselves to null counter-rotating
trajectories in the equatorial plane.

Finally, a third possibility is to try to overspin either the MP black
hole or the fat black ring beyond their maximal value of $j=1$. This
would imply a violation of cosmic censorship.

Thus motivated we study null geodesics in the equatorial plane of a black
ring. The geodesic
equation is simplified using the conserved energy $\delta E$ and $\delta
J$ associated with the Killing vectors
$(\p/\p t)^\mu$ and $(\p/\p\psi)^\mu$, \be
  \delta E = - g_{\mu\nu} (\p/\p t)^\mu \dot{x}^\nu \, ,
  ~~~~~
  \delta J =  \,  g_{\mu\nu} (\p/\p\psi)^\mu \dot{x}^\nu \, .
\ee
The sign in the definition of $\delta J$ is chosen such that
positive $\delta J$ gives a co-rotating geodesic and $\delta J$ is
negative for a counter-rotating geodesic.


\subsection{Black ring geodesics}
\label{geoBR}
We study null geodesics in the plane of the ring, restricted to the
region outside the ring, $x=-1$. From $g_{\mu\nu}
\dot{x}^\mu\dot{x}^\nu=0$ we obtain
\be
  \label{xiy}
  \dot{y}^2  = \xi(y) \equiv A(y,\nu,R) (\delta E)^2+ 2 B(y,\nu,R)
  (\delta J)(\delta E) +D(y,\nu,R) (\delta J)^2 \, ,
\ee
where
\begin{eqnarray*}
A(y,\nu, R)&=& - g^{tt} \,g^{yy} \Big|_{x=-1} 
=-(1+ y)^2 \frac{(1+ y)^4 C ^2
+ (1 - \lambda)^2 G(y)}{R^2 (1-\lambda)^2 F(y)} \, ,\\[2mm]
B(y,\nu, R) &=& \frac{2 \pi}{\Delta \psi}\,
g^{t\psi} \, g^{yy} \Big|_{x=-1} 
=\frac{2 \pi}{\Delta \psi}\frac{(1 +
y)^5 C }{R^3 (1 -
\lambda)^2} \, ,\\[2mm]
D(y,\nu,R) &=& - \left( \frac{2 \pi}{\Delta \psi} \right)^2\,
g^{\psi\psi} \, g^{yy} \Big|_{x=-1} 
=-\frac{4 \pi^2}{\left(\Delta
\psi\right)^2}\frac{(1+y)^4 F(y)}{R^4 (1 - \lambda)^2} \, .
\end{eqnarray*}
The analysis is done for the equilibrium black rings, so it is here
understood that $\lambda = \lambda_c$ as given by \reef{balance}.
The factors of
$\frac{\Delta \psi}{2 \pi}$ are required in order to make the angle
$\psi$ canonically normalized at infinity.

Finding the efficiency coefficient $\alpha$ is the problem of
simultaneously solving $\xi(y)=0$ and
$\xi'(y)=0$. Making a coordinate transformation $y=y(r)$, eq.~\reef{xiy} becomes
$\dot{r}^2=U(r)$, where $U(r) = \xi(y(r))/(y'(r))^2$.
A solution to the problem $U(r)=0=U'(r)$ is also a solution to the
original problem, provided that $1/y'(r)$ does not vanish for the
solution.
Now define
\be
  r = - R \frac{1-\nu}{\nu(1+y)}  \, .
\ee
Asymptotically, $y\to-1$, we have $r\to \infty$, and the normalization
is chosen such that $r=R$ at the horizon $y=-1/\nu$. Next set $\delta J
= \alpha\, \delta E/\Omega$, with $\Omega$ given in \reef{Omega}. We are
assuming that the ring is balanced, so that in all expressions
$\lambda$ is eliminated using \reef{balance}.

It is now easy to see that $U(r)$ has a maximum at $r=r_c$ with
\be
  r_c^2 = \frac{R^2 (1+\nu)(1-2\alpha + \nu)^2}{2 \nu\, (1-\nu)^2} \, .
\ee
Next, $U(r_c)=0$ reduces to
\be
  1+3 \nu^2 - 2\alpha^2 (1+\nu)
  + 2 \sqrt{2 \nu\, (1+\nu)} |1-2\alpha+\nu| = 0 \, .
\ee
There are two solutions,
\be
  \alpha = - \sqrt{\nu} \pm \frac{(1+\sqrt{\nu})^2}{\sqrt{2}\sqrt{1+\nu}} \, ,
\ee
corresponding to co-rotating ($+$) and counter-rotating ($-$) geodesics.
In both cases, $|\alpha| \to 1/\sqrt{2}$ in the limit $\nu \to 0$.

The analysis for the Myers-Perry black hole with a single angular
momentum was done in \cite{EM} (for arbitrary dimension), and can
also be recovered from our results above by taking the limit to the
MP bh as described in \cite{review, RE}. The efficiency coefficient
for MP black holes is
\be
  \alpha = \frac{a}{m} \left( \pm 2 \sqrt{m} -a\right)
\ee
with ``$+$'' for co-rotating and ``$-$'' for counter-rotating
geodesics.

\begin{figure}
\begin{picture}(0,0)(0,0)
\put(85,-1){$\scriptstyle{j}$}
\put(11,45.5){$\scriptstyle{\alpha}$}
\put(8.7,42.5){$\scriptscriptstyle{1}$}
\put(7,30){$\scriptscriptstyle{\frac{1}{\sqrt{2}}}$}
\put(38,-4){$\scriptscriptstyle{\sqrt{\frac{27}{32}}}$}
\put(46.4,-3){$\scriptscriptstyle{1}$}
\put(81,-3){$\scriptscriptstyle{2}$}
\put(24,20){\scriptsize MP}
\put(73,30){\scriptsize BR}
\put(165,-1){$\scriptstyle{j}$}
\put(90,45.5){$\scriptstyle{|\alpha|}$}
\put(88.7,42.5){$\scriptscriptstyle{3}$}
\put(88.7,28.4){$\scriptscriptstyle{2}$}
\put(88.7,13.8){$\scriptscriptstyle{1}$}
\put(86.2,10){$\scriptscriptstyle{\frac{1}{\sqrt{2}}}$}
\put(118.3,-4){$\scriptscriptstyle{\sqrt{\frac{27}{32}}}$}
\put(126.2,-3){$\scriptscriptstyle{1}$}
\put(161,-3){$\scriptscriptstyle{2}$}
\put(108,13){\scriptsize MP}
\put(153,20){\scriptsize BR}
\end{picture}
\centerline{
\includegraphics[width=7cm]{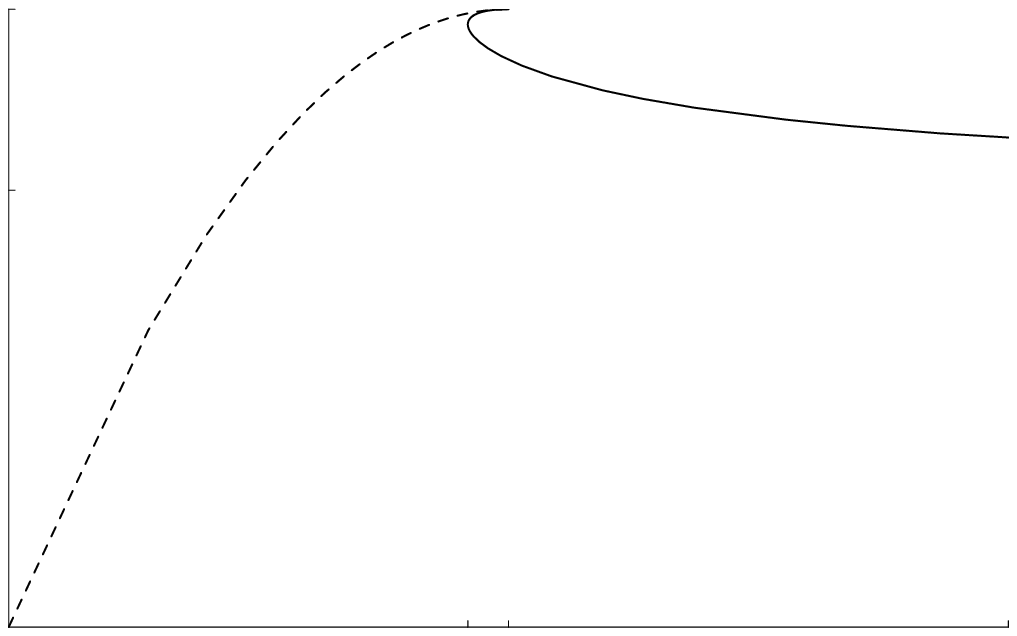}
~~~~~
\includegraphics[width=7cm]{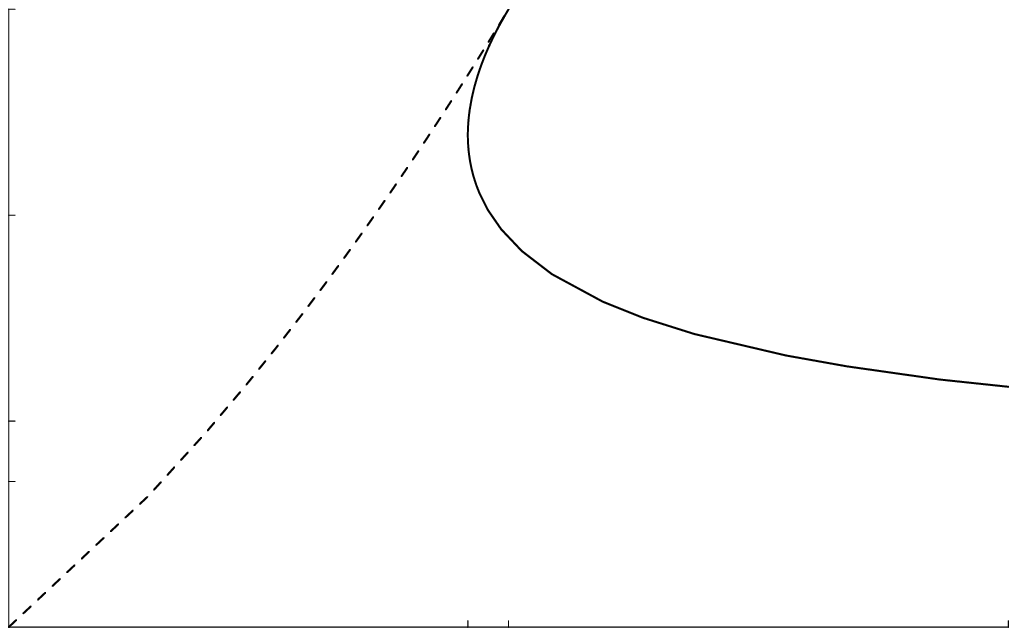}}
\vspace{1mm}
\caption{\small The efficiency parameter $\alpha$ for co-rotating (left) and
counter-rotating (right) null geodesics in the equatorial plane plotted
versus the angular momentum $j$. For the counter-rotating case, $\alpha$
is negative so the plot actually shows $-\alpha$.
The dashed curves shows $\alpha$ for the Myers-Perry black hole, the
full curve is $\alpha$ for the black ring.
}
\label{fig:alpha}
\end{figure}

Figure \ref{fig:alpha} shows $|\alpha|$ versus $j$ for the black ring
and the Myers-Perry black hole for the co-rotating and counter-rotating
null geodesics. It is curious that for thin black rings, the efficiency
of shedding off angular momentum via emission of null particles {\it
decreases} with the angular momentum of the ring, contrary to what
happens for MP black holes. This is related to the property
that, for a thin black ring of fixed mass, the angular velocity
decreases with angular momentum (see figure \ref{fig:double}).

We consider now the four cases of absorbing/emitting
co-(counter-)rotating null particles in the equatorial plane, and the
possible associated instabilities mentioned above (the last fourth case
is included for completeness):

\paragraph{Absorption of co-rotating null particles: Impossibility
of overspinning the MP black hole and fat black ring.}
It is clearly possible to spin up a generic Myers-Perry black hole or
black ring by throwing in a co-rotating null particle: according to
eq.~\reef{deltaS}, this requires $0<\alpha<1$, which holds for both the
MP black hole and the black ring. It is easy to check that in both cases
the effect is that the reduced angular momentum increases, $\delta j >
0$. However, in the singular limit of $j \to 1$ ($\nu \to 1$ for the
fat black ring), one finds that $\delta j \to 0$, so cosmic censorship is
not violated in the process.

\paragraph{Absorption of counter-rotating null particles:
Destabilizing the minimally spinning black ring by slowing it
down.}
The counter-rotating null geodesics have $\alpha<0$, and hence again
$\delta \mathcal{A}_H$ will be positive. It is therefore possible to
decrease the angular momentum, and hence $j$, of the MP black hole or
the black ring by absorption of a null particle. For the black ring with
minimal angular momentum, $j=\sqrt{27/32}$, there is no other black ring
it can evolve into. We expect it to collapse to a Myers-Perry
black hole, which, for that value of $j$, has (much) higher entropy.

\paragraph{Emission of co-rotating null particles: No spontaneous
decay by radiation.}
The spontaneous emission ($\epsilon=-1$) of a null particle with energy
$\delta E$ and $\delta J>0$ requires $\alpha \ge 1$. However, we always
have $\alpha\le 1$ for both the MP black hole and the black ring have,
with equality only for the singular solution with $j=1$ (see figure
\ref{fig:alpha}). So the Myers-Perry black hole and the black ring are
not unstable to the spontaneous emission of null particles carrying away
energy and angular momentum.

\paragraph{Emission of counter-rotating null particles: No
spontaneous spin-ups.}
It is impossible to increase the angular momentum of the black hole by
spontaneously radiating away a null particle in the equatorial plane:
with $\epsilon=-1$ and $\delta J<0$ (so $\alpha<0$) we always have
$\delta \mathcal{A}_H<0$.


\section{Thermodynamics
and Euclidean Black Rings} \label{sec:thermo}

Fig.~\ref{fig:phase} can be regarded as a microcanonical phase diagram,
where the thermodynamic potential is the entropy $S[M,J]$. It is
interesting to study other thermodynamical ensembles, in particular the
grand-canonical ensemble where the variables are the
temperature $T$ and angular velocity $\Omega$ at the horizon. A previous
related study can be found in \cite{ast}.

A frequently used method derives the grand potential $W[\Omega,T]$ from
the saddle-point approximation to the Euclidean black hole partition
function, in terms of the action of a {\it real} Euclidean regular
section of the solution, as $I_{Euc}=W[\Omega,T]/T$ (recall that,
without any other boundary terms, the Einstein-Hilbert-Gibbons-Hawking
action is extremized at fixed $T$ and $\Omega$). For the black ring,
such a real, non-singular Euclidean metric does not seem to exist.

To see this, recall first that for generic spinning black holes the
Euclidean rotation $t\to -i\tau$ does not give a real Euclidean metric,
due to the presence of off-diagonal time-space metric components. To
obtain a real Euclidean solution the angular velocity parameter needs to
be analytically continued to imaginary values. In the case of the Kerr
or Myers-Perry solutions this is achieved by continuing $a\to i a$. In
the parametrization above we would take $0<\lambda<\nu<1$ and
$-1/\nu\leq y\leq - 1$, and then take the limit to the MP black hole.

If we try to do the same for the black ring, we find that the range of
values $0<\lambda<\nu<1$ is incompatible with the balance condition
\reef{balance} and therefore the instanton will possess a naked conical
singularity. Instead we may try to maintain the condition
\reef{balance} in order to eliminate the conical singularity, but then
change the range of values of $\nu$ to make $C$ purely imaginary so as to
obtain a real solution. For real $\nu$ this requires $|\nu|>1$. It is
then easy to see that if the instanton is to be Euclidean, real, and
asymptotically flat, it will have a naked curvature singularity. So it
does not seem possible to construct a non-singular real Euclidean metric
from the black ring.

However, the use of complex Euclidean geometries, where only time and
not the parameters of the solution is Wick-rotated, was advocated in
\cite{BMY}. These solutions can still yield a real Euclidean action, and the
resulting grand potential coincides with the conventional thermodynamics
definition
\be
W[\Omega,T]= M-TS-\Omega J
\ee
where the entropy is given by the area law $S=\frac{\mathcal{A}_H} {4G}$ and
the temperature by the surface gravity $T=\kappa/2\pi$ (see also
\cite{ast}). Using the
Smarr relation $\frac{2}{3} M=TS+\Omega J$,
\be
W[\Omega,T]=\frac{M}{3},
\ee
which is actually valid for both the Myers-Perry black hole and the
black ring.\footnote{It may be worth noting that this same result can be
obtained from a calculation of the Euclidean action of the real solution
obtained by setting $0<\lambda<\nu<1$ and neglecting the contribution to
the action from the conical singularity.}

\begin{figure}
\begin{picture}(0,0)(0,0)
\put(82,-1){$\scriptstyle{T/\Omega}$} \put(0,50){$\scriptstyle{W\,
\Omega^2}$} \put(0,46){$\scriptscriptstyle{\frac{\pi}{8}}$}
\put(40.4,-3){$\scriptscriptstyle{1}$}
\put(78.6,-3){$\scriptscriptstyle{2}$} \put(6.5,10){\scriptsize{MP}}
\put(50,8){\scriptsize{BR}}
\end{picture}
\centering{
\includegraphics[width=3in]{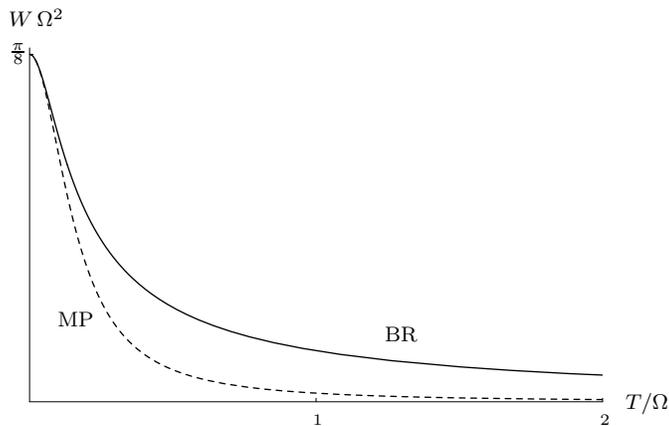}}
\caption{\small Grand-canonical potential $W[T,\Omega]$ as a function of $T$, for fixed
$\Omega$. Black ring is solid curve, MP black hole is dotted curve.}
\label{fig:grandcan}
\end{figure}

Fig.~\ref{fig:grandcan} is a plot of the grand-canonical potential in
which we have fixed the scale by fixing $\Omega$, and then vary $T$.
{\em I.e.,} we plot $\Omega^2 W$ vs.~$T/\Omega$. This shows some
significant differences with respect to the microcanonical ensemble.
There is always one black ring (and only one) and one MP black hole with
the same values of $\Omega$ and $T$. Both kinds of solutions exist for
all possible values of $\Omega$ and $T$. The global thermodynamic
stability in the grand-canonical ensemble is determined by minimizing
the potential $W$: the MP black hole always has lower potential than the
black ring. So in this ensemble the MP black hole is always
thermodynamically favored over the black ring.

Of course it is not unusual that the phase structure and global
thermodynamical stability of black holes depend on the ensemble
considered, since some ensembles possess conserved charges
that others do not. For
the system at hand, other ensembles may be studied, which result in
curious differences. In the canonical ensemble, the free energy
$F[T,J]=M-TS$ exhibits a `swallowtail' structure, but different than the
microcanonical one. Fixing $T$, there is a range of $J$ for which there
are two MP black holes and only one black ring, \ie opposite to the
microcanonical phase diagram. At small $J$ one of the MP black holes
minimizes $F$ and so is preferred, then at a certain value of $J$ we
cross the swallowtail and the black ring is preferred, and for large $J$
only the black ring exists. The `enthalpy' $H[M,\Omega]=M-\Omega J$
gives yet a different phase structure. In this case, the solutions exist
only in a finite range of parameters, and there are always one MP black
hole and one black ring for each value of parameters. But now the
enthalpy is always minimized by the black ring, and so in this ensemble the
black ring is always preferred.


\setcounter{equation}{0}
\section{Discussion}
\label{sec:discussion}

We have found evidence that
\begin{itemize}

\item Fat black rings are unstable to radial perturbations, while thin
black rings are radially stable.

\item Most thin black rings also suffer from GL instabilities.

\item The minimally spinning ring is unstable.

\end{itemize}

As discussed in sec.~\ref{sec:turning}, the first instability result in
this list (although not the nature of the instability) actually follows
from general considerations studied in \cite{loar}. The possible
appearance or disappearance of unstable modes along the curve of
equilibrium states comes from the existence of turning points or
bifurcations, and the current phase diagram allows us to identify only
the cusp as a point where unstable modes appear. It must be noted,
though, that it is possible that only axisymmetric modes are accounted
for by the Poincar\'e method \cite{loar}. In this case the GL
instability would likely be invisible to it, and it can possibly switch
off at some value of $j$ along the thin black ring branch. This would
allow the existence of absolutely stable thin black rings for moderate
values of $j$, perhaps within a range roughly coincident with the range
where non-uniqueness happens. Evidently this requires the absence of any
unstable modes besides those we have studied. This problem clearly
deserves further study.

It does not seem probable that additional neutral black hole solutions
with a single connected horizon and as many as three Killing symmetries
exist in $D=5$. On the other hand, black hole phases with broken axial
symmetry along $\phi$ have been conjectured \cite{harvey}. They may play
a role if the MP black hole actually becomes unstable near $j=1$.

The conclusions above apply to neutral black rings with a single angular
momentum. The addition of a second angular momentum along the $S^2$ may
change the behavior of some of the instabilities we have analyzed.
However, doubly-spinning black rings are likely to suffer from
superradiant instabilities peculiar to extended black objects
with a rotating sphere \cite{odias}.

It is therefore likely that neutral black rings, with one or two angular
momenta, are dynamically unstable to some kind of perturbation over a
wide range of parameters. If they are actually unstable for all
parameter values, then black rings may be difficult, perhaps impossible,
to produce in physical processes. However, does this imply that these
black rings are irrelevant as physical objects?

The answer, of course, depends on the relative value between the time
scale for the evolution of the instabilities and the time scales of the
physical processes considered. Slow instabilities may become essentially
negligible for many purposes.

The information we have at the moment about the time scales for
instabilities is rather scarce. We have argued in
sec.~\ref{sec:frag} that the time scale for the Gregory-Laflamme
instability of thin rings is very quick, eq.~\reef{tgl}, although it
could turn into a much slower one, eq.~\reef{tgw}, if the GL instability
of black strings slows down, and possibly halts, as suggested in
\cite{HM}. The superradiant instabilities of black rings with two
angular momenta are expected to be very slow too \cite{odias}.

As for the time scale for radial instabilities, we do not really know.
Naively, the curvature $V''$ of the radial potential around the
equilibrium point gives the square of the imaginary frequency of the
unstable mode. However, it is not quite clear whether the potential
\reef{vofr} (even if properly normalized) can be used for this purpose,
since the time evolution of the system presumably happens through
highly-dynamical geometries that have little to do with the
conically-deformed ones that we have studied.

Given the uncertainty about many dynamical aspects of higher-dimensional
gravity, one should also be cautious when speculating about the
evolution of these instabilities. It seems natural
to expect that the radial instabilities of fat black rings follow the
evolution suggested by the potential in fig.~\ref{fig:VofR}.
Less clear is the evolution of the GL instability. The break-up
into black holes flying apart gives a simple physical picture. However,
it would not be unprecedented in this field if wholly new phenomena were
at play.

The inclusion of charges does improve stability: supersymmetric black
rings \cite{susy1} are expected to be stable, and near-supersymmetric ones
\cite{NSring} probably too. There also exist black rings with dipole
charges but no conserved charges \cite{RE}. We have performed the
analysis of radial stability for these dipole rings (see appendix
\ref{app:dipolerad}), and the conclusions are similar to what we have
found for neutral rings: when the dipole charge is fixed, in addition to
the area and spin, thin rings appear to be radially stable, fat rings
unstable. This also applies, in particular, for the extremal dipole
rings (which are not supersymmetric) with a regular horizon. One expects
that close to the extremal limit the GL instability switches off. So it
is very well possible that thin dipole rings close to extremality are
dynamically stable.

It might be possible to gain more insight on the dynamics of
higher-dimensional black holes using their string theory description.
Supersymmetric and near-supersymmetric black holes can be understood as
excited D-brane systems in string theory. D-branes are highly dynamical
objects that can move and ripple, so from the point of view of string
theory it comes as no surprise that black holes are also dynamical
objects.
Picturing
neutral black holes as brane-antibrane systems \cite{HMS, DGK} --- the
branes and antibranes entering with equal and opposite charges thus
leaving no net charge (nor any higher moments of charge) --- there are
even microscopic indications of phenomena such as the GL instability
\cite{DGK}. String models for black rings with dipole charges have been
suggested \cite{RE}. In these models, strings (or branes) wind around
the circle of the ring. With this in mind it is perhaps noteworthy that
at leading order we have found the balance of (thin) black rings to be
dominated by a string-like tension rather than a gravitational force.
A better understanding of the microscopics of black rings may therefore
give valuable hints about their macroscopic dynamics.


\section*{Acknowledgements} We are grateful to Don Marolf for
conversations and suggestions. RE is grateful to Harvey Reall and
Tetsuya Shiromizu for early collaboration and discussion of several of
the issues addressed in this paper. AV is thankful to Gautam Sengupta
for guidance during an undergraduate project on black rings. We are
particularly indebted to Veronika Hubeny for making us realize a crucial
mistake in an earlier version of this paper. We also thank the anonymous
referee for pointing us to the arguments in Sec.~\ref{sec:turning}. We
would like to thank the Kavli Institute for Theoretical Physics at UC
Santa Barbara for hospitality, support and a stimulating program
``Scanning New Horizons: GR Beyond 4 Dimensions'' during which this work
was initiated. HE would also like to thank the Niels Bohr Institute for
hospitality during the final stages of the project. RE is supported in
part by DURSI 2005 SGR 00082, CICYT FPA 2004-04582-C02-02 and EC FP6
program MRTN-CT-2004-005104. HE was supported by a Pappalardo Fellowship
in Physics at MIT and by the US Department of Energy through cooperative
research agreement DF-FC02-94ER40818. AV was supported in part by NSF
under Grant No.~PHY03-54978 and by funds from the University of
California. This research was supported in part by the National Science
Foundation under Grant No.~PHY99-07949.


\appendix
\setcounter{equation}{0}
\section{Isometric Embedding}
\label{app:embed}
The idea of isometric embedding is to visualize a curved geometry in
flat Euclidean space in a way that preserves the distances. Not all 2d
surfaces can be embedded isometrically in 3d euclidean space, but the
two-sphere of the black ring can be embedded in $\mathbb{E}^3$.

Consider a surface H with metric
\be
  ds_\rom{H}^2 = g_{\theta\theta} \, d\theta^2
  +g_{\phi\phi} \, d\phi^2 \, ,
\ee
where the metric functions depend on $\theta$ only and $0\le \theta
\le\pi$. We wish to embed this into a non-physical 3d euclidean space
\be
  ds^2(\mathbb{E}^3) = dx^2 + dy^2 + dz^2 \, .
\ee
Let
\be
  x = f(\theta) \, \cos\phi \, , ~~~~~~
  y = f(\theta) \, \sin\phi \, , ~~~~~~
  z = g(\theta) \, .
\ee
Then from $ds_\rom{H}^2 = ds^2(\mathbb{E}^3)$ we get
\be
  f = \sqrt{g_{\phi\phi}}\, , ~~~~~~~
  {g'}^2 = g_{\theta\theta} - {f'}^2  \, .
\ee
The necessary condition for the embedding of the surface H into
$\mathbb{E}^3$ is then that
\be
  \label{embed}
  g_{\theta\theta} \ge \left( \partial_\theta \sqrt{g_{\phi\phi}}\right)^2 \, .
\ee
The embedding condition \reef{embed} is satisfied for the two-sphere
of the black ring horizon. Fig.~\ref{fig:S2dist} shows the
cross-section of the black ring two-sphere for various values of $j$
($\nu$) with the mass $G M$ fixed. The figure is produced as a
parametric plot of $f(\theta)$ versus $g(\theta)$, suppressing the
azimuthal angle $\phi$. Fig.~\ref{fig:ringsillustrated} was made using
the isometric embedding of the $S^2$ cross section of
fig.~\ref{fig:S2dist}, but with the $S^1$ fixed to match the inner
radius of the $S^1$. Thus fig.~\ref{fig:ringsillustrated} is simply a
cartoon of the black ring geometry.


\setcounter{equation}{0}
\section{Ring Radii and Rotation Velocity}
\label{app:radii}

Since in general the black ring horizon is distorted, the definitions of
the radii of the circle and the two-sphere are ambiguous. In the
following, we introduce different definitions. For definitions of the
$S^2$ radii, we shall assume the balancing condition
\reef{balance}, but in applications is useful to also define
the $S^1$ radii for black rings away from mechanical
equilibrium.

\subsection{Radius of $S^2$: $R_2$}

One possibility is to define the radius $R_2^\circ$ at the `equator'
where the sphere is fattest, namely at
\be\label{fatxo}
  x_\circ = \frac{-1+\sqrt{1-\nu^2}}{\nu}  \, ,
\ee
as can be seen from the horizon metric \reef{horizon}. This gives
the `equator radius' of the $S^2$,
\be
  R_2^\circ = R\, \sqrt{\frac{2-2 \sqrt{1-\nu^2} }{1+\nu^2}}\, .
\ee
Another option is to use the area of the $S^2$ on the horizon (at
constant $t$ and $\psi$) to define the radius as
\be
  R_2^\rom{area} =\sqrt{\frac{\mathcal{A}_{S^2}}{4\pi}}=
  \left(
  \frac{1}{2}\frac{\Delta\phi}{2\pi}
   \int_{-1}^1 dx\,\sqrt{g_{xx}\, g_{\phi\phi}} \Big|_{y=-1/\nu}
  \right)^{1/2} \ .
\ee
The definitions $R_2^\circ$ and $R_2^\rom{area}$ agree well for black
rings on the thin ring branch, as can be seen in
fig.~\ref{fig:S2radii}. The agreements is expected, since the $S^2$ is
nearly round for thin black rings (see fig.~\ref{fig:S2dist}).
As $\nu\to 0$ both $R_2^\circ$ and $R_2^\rom{area}$ tend to the value
$R \nu$ equal to the area radius of the $S^2$ for the boosted black
string.

In the singular limit $\nu\to 1$, both $R_{2}^\circ$ and
$R_2^\rom{area}$ go to zero.
The effect is that the horizon flattens out before finally
disappearing for $\nu=1$.

\subsection{Radius of $S^1$: $R_1$}
\label{sec:r1}

The radius of the circle measured at a
given polar angle on the $S^2$, specified by $x$, is
\beq\label{r1x}
  R_1^x=\frac{\Delta\psi}{2\pi}\sqrt{g_{\psi\psi}(y=-\nu^{-1},x)}
 =R\,\sqrt{\frac{\lambda(1+\lambda)}{\nu(1+\lambda x)}}\,.
\eeq
For very thin rings, $\lambda, \nu \to 0$, the $S^1$ radius goes to
$R\sqrt{\lambda/\nu}$ independently of $x$. We will see below that the
factor $\sqrt{\lambda/\nu}$ is a relativistic boost effect.

There are several values of $x$ of particular interest:
inside of the ring at $x=1$, on the outside at $x=-1$, or at
the $S^2$ `equator' at $x=x_\circ$. Assuming, for illustration, the
condition for equilibrium
\reef{balance} one finds
\be
  R_1^\rom{inner} =R\, \sqrt{\frac{2}{1+\nu^2}}\,  ,
  ~~~~~
  R_1^\rom{outer} =R\, \frac{1+\nu}{1-\nu} \sqrt{\frac{2}{1+\nu^2}} \, ,
  ~~~~~
  R_1^\circ =R\,\frac {\sqrt{2} \, (1+\nu)}
   {\sqrt{(1+\nu^2) \big(2 \sqrt{1-\nu^2} + \nu^2 -1\big) }} \, .
  \label{S1radii}
\ee
Another useful measure for the $R_1$ is based on the horizon area
$\mathcal{A}_\rom{H}$:
 \be
  R_1^\rom{area} = \frac{\mathcal{A}_\rom{H}}{8 \pi^2 (R_2^\circ)^2} \, .
\ee
For $\nu \to 1$, $R_1^\rom{inner}$ and $R_1^\circ$ go to
zero while $R_1^\rom{outer}$ and $R_1^\rom{area}$
diverge: as the black
ring approaches the singularity, the hole inside
the ring shrinks to zero size while the outer radius goes to
infinity.

\begin{figure}[t!]
\begin{picture}(0,0)(0,0)
\put(113,50){$\scriptstyle{R_1}$} \put(164,-1){$\scriptstyle{j}$}
\put(84,-4){$\scriptscriptstyle{\sqrt{\frac{27}{32}}}$}
\put(114,-3){$\scriptscriptstyle{1}$}
\put(145.5,-3){$\scriptscriptstyle{1.1}$}
\put(116,8.8){$\scriptscriptstyle{1}$}
\put(116,18.4){$\scriptscriptstyle{2}$}
\put(116,27.5){$\scriptscriptstyle{3}$}
\put(116,36.6){$\scriptscriptstyle{4}$}
\put(138,21){$\scriptstyle{R^\rom{outer}_1}$}
\put(141,10){$\scriptstyle{R^\rom{inner}_1}$}
\put(104,22){$\scriptstyle{R^\rom{area}_1}$}
\put(109,5){$\scriptstyle{R^\circ_1}$} 
\put(36,50){$\scriptstyle{R_2}$} \put(79.5,-1){$\scriptstyle{j}$}
\put(0,-4){$\scriptscriptstyle{\sqrt{\frac{27}{32}}}$}
\put(36.6,-3){$\scriptscriptstyle{1}$}
\put(75.5,-3){$\scriptscriptstyle{1.1}$}
\put(38.5,14.9){$\scriptscriptstyle{0.1}$}
\put(38.5,38){$\scriptscriptstyle{0.25}$}
\put(2,45){$\scriptstyle{R^{A_\rom{H}}_2}$}
\put(27,37.5){$\scriptstyle{R^\rom{area}_2}$}
\put(25,19){$\scriptstyle{R^\circ_2}$}
\end{picture}
\vspace{1mm}
  \centering{
    \subfigure[
    Radii for the $S^2$.]
    {\label{fig:S2radii}\includegraphics[width=7.5cm]{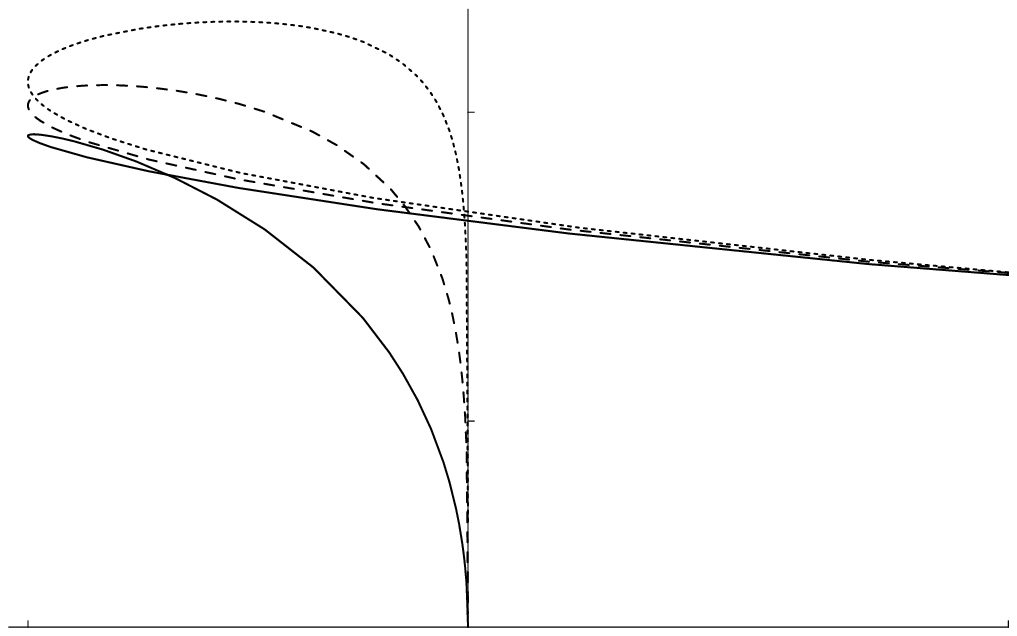}}
    \hspace{6mm}
    \subfigure[
    Radii for the $S^1$.
    ]
    {\label{fig:S1radii}
    \includegraphics[width=7.5cm]{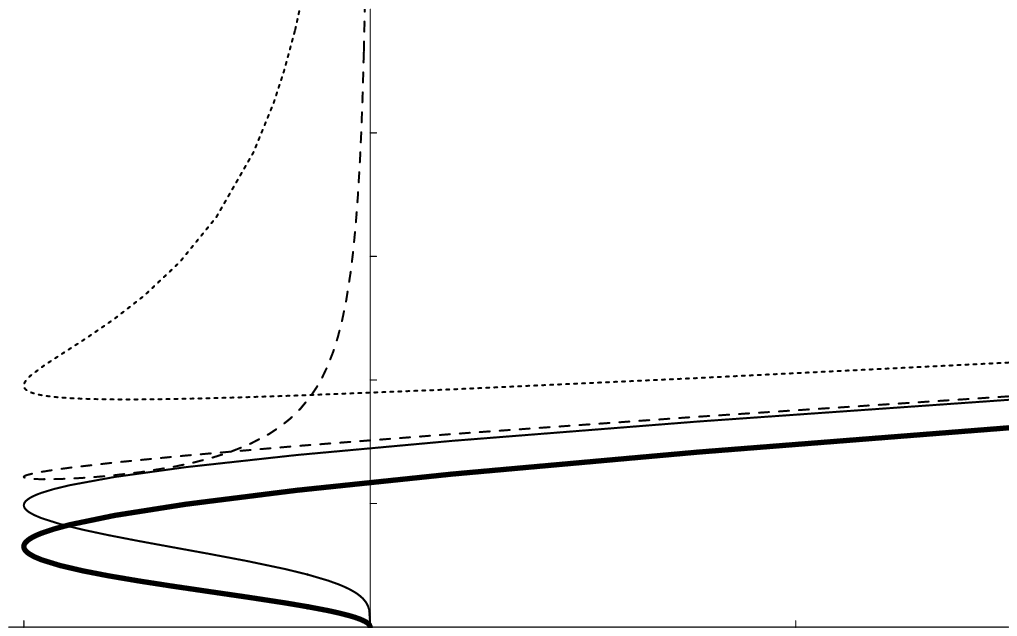}}
    \caption{\small\label{fig:S1S2radii}For fixed mass, $GM=1$, the various
    definitions of radii of the two-sphere (figure \ref{fig:S2radii})
    and of the circle (figure \ref{fig:S1radii}) are plotted vs.~the
    reduced angular momentum $j$.}}
\end{figure}

Figure \ref{fig:S1radii} compares for fixed mass the four definitions
of the circle
radii plotted as functions of the reduced angular momentum $j$.  For
black rings on the thin ring branch ($\nu < 1/2$) the area-radius
and $R_1^\circ$ agree fairly well, especially for large angular momenta.
In fact, at large angular momenta one finds that, for fixed mass, the
$S^1$ radius grows linearly with $j$.

\vspace{2mm}
We can also define an $S^2$ radius $R_2^{\mathcal{A}_\rom{H}}$ based
on the total area $\mathcal{A}_\rom{H}$ of the black ring.
Using $R_1^\circ$ for the radius of the circle to
extract $R_2^{\mathcal{A}_\rom{H}}$ from the total area
$\mathcal{A}_\rom{H}$, this definition agrees well with the other
two definitions for thin black rings, see figure \ref{fig:S2radii}.

\subsection{Rotation velocity}
\label{app:lorentz}

Since the black ring has angular momentum, it experiences effects of
Lorentz contraction. For our purposes in section \ref{sec:gl} it is
desirable to have a way to account for this effect. This requires making
some judicious estimate for the boost velocity $v$ of the black ring.

In the limit of infinite radius the balanced black ring becomes a
boosted black string with velocity $v=\sqrt{1-\nu/\lambda}$ \cite{RE}.
So we require that our definition of $v$ for the black ring reproduces
this limit for $\lambda,\nu \to 0$. It is natural to base the definition
of $v$ on the angular velocity $\Omega$ of the black ring. Indeed for
the boosted black string, $v = \Omega\, R_1(\infty)$, where
$R_1(\infty)$ is the radius of the circle at infinity. For the
black ring there is of course no circle at infinity, so we have to base the
definition on the radius of the $S^1$ of the horizon. Note that for the
boosted black string, the radius of the $S^1$ of the horizon is
$R_1(\rom{horizon})=R_1(\infty)/\sqrt{1-v^2}$.
Thus motivated we define $v$ for the black ring by
\be v = \Omega R_1 \sqrt{1-v^2} \ee \ie
\begin{equation}
\label{velocity} v = \frac{\Omega R_1}{\sqrt{1+(\Omega
R_1)^2}}\, ,
\end{equation}
for some choice $R_1$ of the $S^1$ radius.
If we use $R_1=R_1^x$, then \reef{r1x} gives
\be
v =
\sqrt{\left(1-\frac{\nu}{\lambda}\right)\frac{1}{1+\nu x}} \,.
\ee
Note that we recover
$v=\sqrt{1-\nu/\lambda}$ as $\lambda,\nu\to 0$.


\setcounter{equation}{0}
\section{Radial stability of dipole rings}
\label{app:dipolerad}

The analysis of radial stability in sec.~\ref{sec:radial-general} can be
extended to black rings with dipoles. The same result is obtained as for
neutral rings: fat black rings are radially unstable, thin black rings
are radially stable. That the minimally spinning ring separates the two
stability regimes follows from a general argument as in
sec.~\ref{sec:turning}. The general analysis is rather tedious, so for
the purpose of illustration here we only provide details for a case of
particular significance: dipole rings with extremal (maximal) dipole
charge, and with a regular horizon.

These are black ring solutions of non-dilatonic Einstein-Maxwell theory
(with, possibly, an additional Chern-Simons term, as required by minimal
5D supergravity) with a dipole $q$ \cite{RE}. They can be regarded as
intersections of three stacks of M5-branes over a ring, with equal
numbers of branes in each stack. In the extremal limit the horizon is
degenerate (zero-temperature) but still is non-singular and has finite
area. 

Fixing the scale, the solutions are characterized by a single
dimensionless parameter $\mu$, in terms of which we can express the
reduced spin, area, and dipole as \cite{RE}\footnote{Here $\hat q$ corresponds to $q$
in \cite{RE}. Following \cite{susy1}, we reserve $q$ for the dipole.}
\beq
j^2=\frac{(1+\mu)^6(3+\mu^2)^2}{128\mu(1+\mu^2)^3}\,,
\quad
{a_H}=\sqrt{\frac{\mu(1-
\mu^2)^3(3+\mu^2)}{4(1+\mu^2)^3}}\,,
\quad 
{\hat q}=\sqrt{\frac{\mu(1-\mu)^2}{\pi(1+\mu^2)}}\,,
\eeq
which are valid for the equilibrium configurations. The spin $j$ reaches
a minimum value at the real root of
\beq\label{muc}
\mu_c(3+\mu_c+\mu_c^2)=1\,,
\eeq
(\ie $\mu_c\simeq 0.296$) which separates the branches of thin rings,
$0\leq\mu<\mu_c$, and fat rings $\mu_c<\mu<1$.

Following the analysis in sec.~\ref{sec:radial-general}, we now perturb
these configurations by varying the inner radius
$R_1^\mathrm{inner}$ away from its equilibrium value. This
generates a membrane disk with tension
\beq
\tau=\frac{3}{8G}\left(1-\sqrt{\frac{1-
\lambda}{1+\lambda}}\left(\frac{1+\mu}{1-\mu}\right)^{3/2}\right)
\eeq
where, at equilibrium, $\lambda=\mu(3+\mu^2)/(1+3\mu^2)$.

We have to choose whether we keep fixed the spin and the area, or the
spin and the dipole. We have checked both cases and found analytically
that they yield the same conclusions for radial stability. In
particular, if we fix the spin and dipole we find
\beq
\left(-\frac{d\tau}{d
r}\right)_{J,q}=\frac{9}{4G} \frac{1-
\mu(3+\mu+\mu^2)}{\big(5+\mu(3+3\mu+\mu^2)\big)r}\,.
\eeq
This is positive, hence stable, for thin rings, and as expected, changes
sign precisely at the cusp $\mu=\mu_c$, \reef{muc}, rendering fat rings
radially unstable.

For the non-extremal case, the calculation of
$\left(d\tau/dr\right)_{J,q,\ast}$ with $\ast=M,\mathcal{A}_H$, is
straightforward, and one gets a polynomial in $\nu$ and $\mu$. The only
complication is finding out, for fixed $q$, at which value of $\nu$ is
the cusp separating the branches located. We have done this numerically,
and checked again that fat rings are radially unstable, with stability
changing at the cusp. 

Finally, the results can also be extended to dilatonic dipole rings.


\end{document}